\documentclass[prl,aps,twocolumn,epsfig,10pt]{revtex4-2}
\usepackage{graphicx}
\usepackage{dcolumn}
\usepackage{bm}
\usepackage{lipsum}
\usepackage{xcolor}
\usepackage{soul}
\usepackage{ulem}
\makeatletter

\newcommand{\Rmnum}[1]{\expandafter\@slowromancap\romannumeral #1@}
\makeatother
\begin{document}

\title {Emergent inhomogeneity and non-locality in a graphene field-effect transistor on a near-parallel moir\'{e} superlattice of transition metal dichalcogenides}
\author{Shaili Sett$^{1*}$, Rahul Debnath$^{1*}$, Arup Singha$^{1}$, Shinjan Mandal$^{2}$, Jyothsna K. M.$^{3}$, Monika Bhakar$^{4}$,  Kenji Watanabe$^{5}$, Takashi Taniguchi$^{6}$, Varun Raghunathan$^{3}$, Goutam Sheet$^{4}$, Manish Jain$^{2}$ and Arindam Ghosh$^{1,7}$}
\vspace{1.5cm}
\affiliation{$^1$Department of Physics, Indian Institute of Science, Bangalore 560012, India.}
\affiliation{$^2$Centre for Condensed Matter Theory, Department of Physics, Indian Institute of Science, Bangalore 560012, India.}
\affiliation{$^3$Department of Electrical Communication Engineering, Indian Institute of Science, Bangalore 560012, India.}
\affiliation{$^4$Department of Physics, Indian Institute of Science Education and Research Mohali, Punjab 140306, India.}
\affiliation{$^5$Research Center for Electronic and Optical Materials, National Institute for Materials Science, 1-1 Namiki, Tsukuba 305-0044, Japan.}
\affiliation{$^6$Research Center for Materials Nanoarchitectonics, National Institute for Materials Science,  1-1 Namiki, Tsukuba 305-0044, Japan.}
\affiliation{$^7$Centre for Nanoscience and Engineering, Indian Institute of Science, Bangalore 560012, India.} 
\affiliation{$^*$ Equal contribution from both authors}

\begin{abstract}
At near-parallel orientation, twisted bilayer of transition metal dichalcogenides exhibit inter-layer charge transfer-driven out-of-plane ferroelectricity that may lead to unique electronic device architectures. Here we report detailed electrical transport in a dual-gated graphene field-effect transistor placed on 3R stacked twisted bilayer of WSe$_2$ at a twist angle of 2.1$^\circ$. We observe hysteretic transfer characteristics and an emergent charge inhomogeneity with multiple local Dirac points as the electric displacement field ($D$) is increased. Concomitantly, we also observe a strong non-local voltage signal at $D \sim 0$ V/nm that decreases rapidly with increasing $D$. A linear scaling of the non-local signal with longitudinal resistance suggests edge mode transport, which we attribute to the breaking of valley symmetry of the graphene channel due to the spatially fluctuating electric field from the moir\'{e} domains of the underlying twisted WSe$_2$. A quantitative analysis connecting the non-locality and channel inhomogeneity suggests emergence of finite-size domains in the graphene channel that modulate the charge and the valley currents simultaneously. This work underlines efficient control and impact of interfacial ferroelectricity that can trigger a new genre of devices for twistronic applications.

\end{abstract}

\maketitle

\section{Introduction}
 Ferroelectricity in two dimensional (2D) materials is an emerging area of research with only a handful of 2D ferro-layers identified till date \cite{xue2021emerging}. From a technological point of view, 2D ferroelectrics show a strong potential for novel device architectures in multiple applications \cite{ryu2020empowering, kim2023wurtzite}. Ferroelectric field-effect transistors (FeFETs) based on conventional 3D ferroelectrics/2D semiconductors and other 2D ferroelectrics (like CuInP$_2$S$_6$ in multilayers of thickness $\approx$ 90 nm \cite{li2022electrostatic}) have been explored for non-volatile memory devices, neuromorphic computing \cite{covi2022ferroelectric}, opto-synaptic applications \cite{liu2023ferroelectric} and other versatile functions \cite{wang2023towards}. In this context, there is new interest on engineering ferroelectricity in van der Waals (vdW) materials, achieved by intentionally disrupting the centro-symmetry of an even number of 2D layers through rotational alignment \cite{zhang2023ferroelectric, zhang2023twisted}. The presence of a clean interface between layers, small depolarization fields, and the atomic layer thickness of 2D twisted bilayers offer an alternative for overcoming interfacial challenges and fundamental limitations imposed by the vertical size scaling in thin films of conventional ferroelectrics like doped PbTiO$_3$ and hafnia \cite{xue2021emerging,wang2023towards}. 

 The atomic registry of 2D bilayers in 3R stacking configuration breaks the inversion symmetry at the interface, creating an out-of-plane electric dipole moment and enabling interfacial charge transfer between the two layers. The resultant moir\'{e} domains of opposite polarization lie adjacent to each other, forming an anti-ferroelectric multi-domain structure in the moir\'{e} domain length scales. A distinct feature of polarization switching in stacking-engineered 2D layers compared to their 3D counterparts is the remarkable phenomenon of lateral shifts mediated by domain wall motion \cite{wang2023towards}. Ferroelectricity, achieved via this mechanism, has been recently explored in a wide range of layered materials ranging from Bernal-stacked bilayer graphene and hBN \cite{zheng2020unconventional}, AB or BA stacked twisted hBN \cite{yasuda2021stacking} and near-parallel twisted transition metal dichalcogenides (TMDCs) like WSe$_2$, MoSe$_2$, MoS$_2$ etc \cite{wang2022interfacial, weston2022interfacial}.
 
 \begin{figure*} 
\includegraphics[width=1\linewidth]{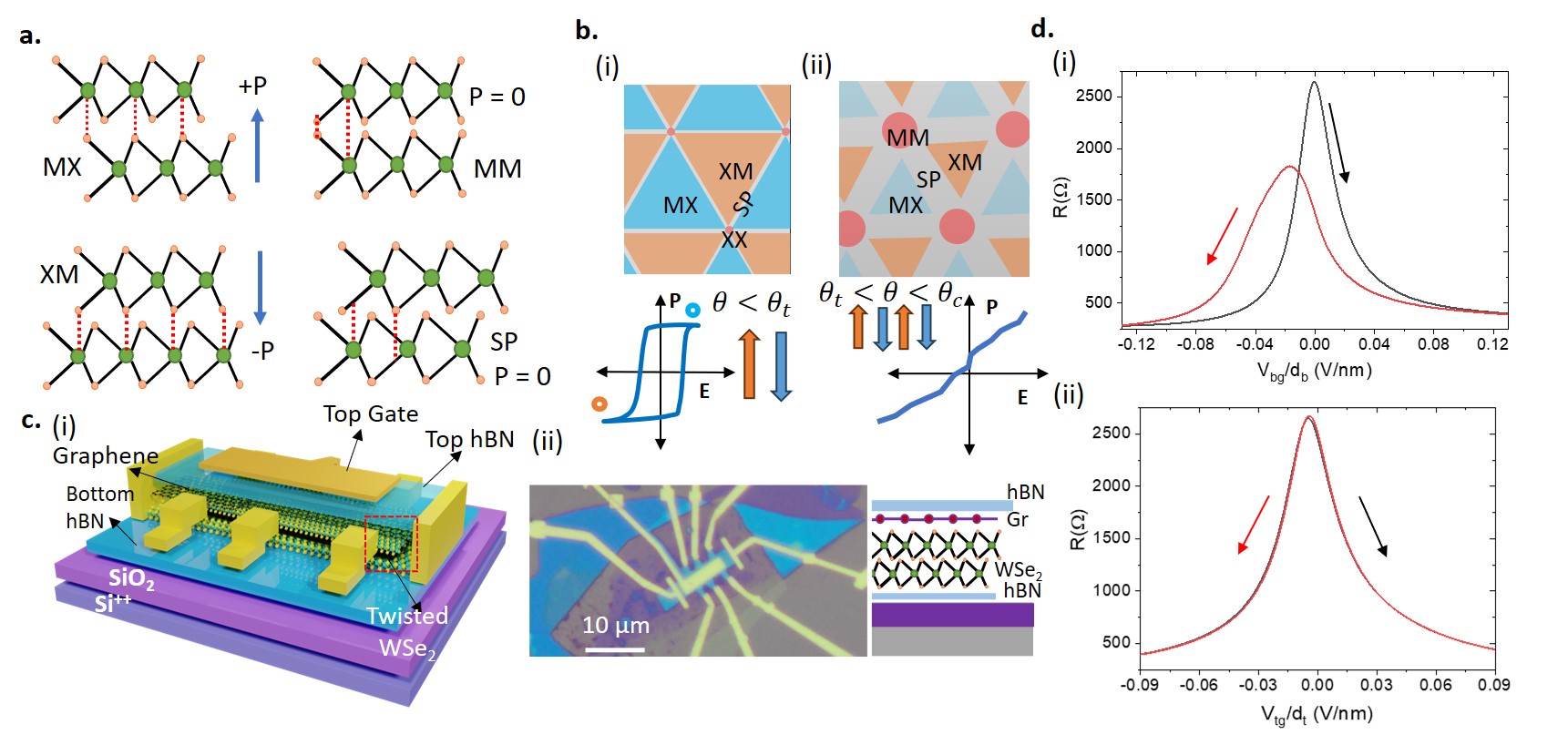} 
\caption{(a) Atomic arrangement of high symmetry stackings; MX (metal on chalcogen), XM (chalcogen on metal), MM (metal on metal); random stacking corresponding to the domain wall that has a saddle point (SP) in energy. The out-of-plane polarization for these stackings are indicated by P. (b) Schematic of a moir\'{e} superlattice for (i) $\theta < \theta_{t}$, which shows ferroelectric state with the conventional P-E loop showing two polarization states. (ii) When $\theta_{t} < \theta > \theta_{c}$, the moir\'{e} superlattice is anti-ferroelectric, with multiple polarization states giving rise to step-like P-E characteristics. (c) (i) Schematic of the cross-sectional view of the hBN-encapsulated Gr/tWSe$_2$ hybrid showing the device architecture. (ii) Optical image of the fabricated device (twist angle $\sim2.1^\circ$). The moir\'{e}  wavelength is $\ell_m = \frac{a}{2\sin(\theta/2)}$, where a = 0.328 nm is the lattice constant of WSe$_2$ and $\ell_m \simeq$ 9 nm for our device. Arrangement of the individual 2D layers shown alongside. (d) Gr/tWSe$_2$ electrical characteristics at room temperature in the absence of an electric field. (i) $R-V_{bg}$ curves for forward (black) and backward (red) scans directions showing hysteresis. (ii)  $R-V_{tg}$ curve in the forward and backward scan directions overlap and have no hysteresis.}
\label{transport1}
\end{figure*}

 At small angles ($\theta$) of twist between the two layers, the moir\'{e} lattice reconstructs significantly to maximize the domain area of the high symmetry stacking regions with the lowest energy \cite{mcgilly2020visualization}. In 3R stacked TMDCs, metal on chalcogen (MX) and chalcogen on metal (XM) stackings maximize to form alternate triangular domains, while the highest energy stacking of metal on metal (MM) occupies the minimum area and appears as a node to the domain wall network (DWN). It is formed by arbitrary stacking order corresponding to a translation from MX domain to XM domain and vice versa \cite{naik2018ultraflatbands, naik2020origin, cazeaux2023relaxation} as shown in Fig.~\ref{transport1}(a). The domain walls are a saddle point (SP) in the energy landscape of the moir\'{e} superlattice and have no out-of-plane polarization component while the MX and XM domains have an out-of-plane dipole moment, due to interlayer charge transfer, and are of opposite sign. Recent studies \cite{bennett2023polar} show that the DWN may be topological in nature with a finite winding number and can possess an in-plane polarization. Fig.~\ref{transport1}(b) shows a schematic of how the different stacking regions evolve with twist angle. The stacking area of each domain, domain wall width and moir\'{e} period is significantly modified, till the two layers are uncoupled (see Supporting Information (SI), section II for atomistic calculations on stacking area versus twist angle dependence). Hence, the ferroelectric polarization properties of the moir\'{e} superlattice can be modified by altering the twist angle between the two layers. In-situ transmission electron microscopy demonstrates that WSe$_2$ bilayers can be tuned to transition from a ferroelectric, to anti-ferroelectric and the usual dielectric \cite{ko2023operando} with increasing twist angle ranges. From $0^\circ$ to a transition angle ($\theta_{t}$  $\le$ 1$^\circ$), the layers can be tuned to become completely ferroelectric, where the moir\'{e} domain size is similar to the system size. On further increase in misorientation, the system becomes anti-ferroelectric up to a critical angle, $\theta_{c}$ $\approx$ 2.1$^\circ$, with a variably small percentage of oppositely polarized domains remaining even at large electric fields (see Fig.~\ref{transport1}(b)). Finally, for twist angles greater than $\theta_{c}$, the layers are decoupled and there is hardly any lattice reconstruction leading to the usual dielectric nature of the bilayers. The regime $\theta \sim \theta_c$ is unique because the length scale of the polarization fluctuations across successive moir\'{e} domains approach $\sim$ few lattice spacings, and thereby potentially impact the valley symmetry of the graphene channel.
 
 Here we investigate electrical transport in vdW heterostructures, with graphene placed over twisted TMDC bilayers, where the TMDC layers are twisted at $\theta \approx 2.1^\circ$. Graphene-TMDC heterostructures have been extensively used to probe the electronic \cite{bhowmik2022spin, sett2021anomalous}, opto-electronic \cite{roy2018number} and spintronic properties \cite{tiwari2020observation}. In the present work, graphene is used as the channel layer placed directly on a near-parallel ($\theta$ = 2.1$^{\circ}$) twisted WSe$_2$ bilayer, thus making the latter a part of the (bottom) gate stack. We observe of an emergent charge inhomogeneity that is manifested in electrical transport at large transverse electric fields. This is accompanied by a giant non-locality, indicating a possibility of breaking the sub-lattice symmetry of graphene, imparted by a spatially fluctuating local electric field from the ferroelectric coupling of the twisted WSe$_2$ bilayer with graphene. 

\begin{figure*} 
\includegraphics[width=1\linewidth]{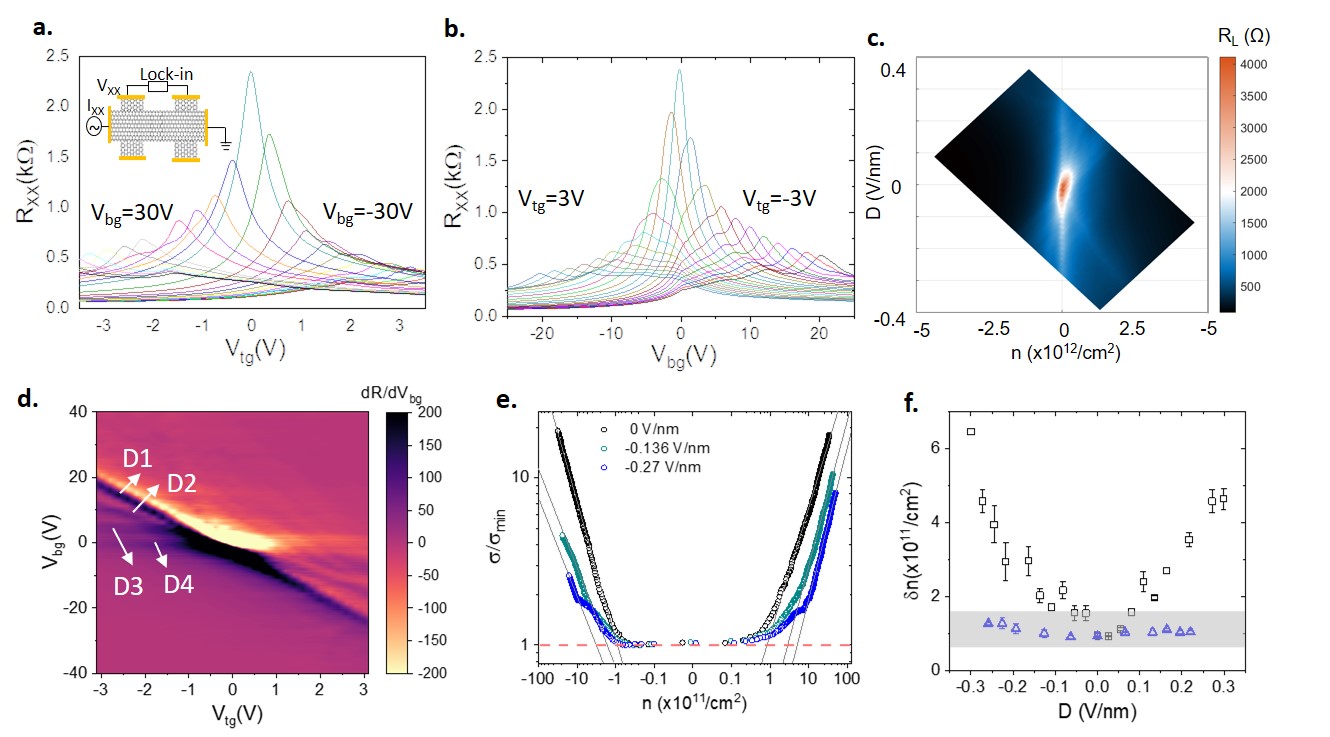} 
\caption{Effect of a displacement field in Gr/tWSe$_2$. (a) $R-V_{tg}$ sweeps at fixed values of $V_{bg}$ ranging from -30 V to 30 V, in steps of 2.5 V at 80 K. (b) $R-V_{bg}$ sweeps at fixed values of $V_{tg}$ ranging from -3 V to 3 V, in steps of 0.3 V at 80 K. (c)  Color map of $R_L$ as a function of total carrier density $n$ and the displacement field $D$ at 0.3 K. (d) The phase space plot (with transresistance in color) at 0.3 K, showing the emergence of multiple local Dirac points D1, D2, D3 and D4, that appear as straight lines diverging from a common origin. (e) Conductivity ($\sigma$) normalized by the minimum conductivity ($\sigma_{min}$) as a function of number density ($n$) in Gr/tWSe$_2$ at $D$ = 0 V/nm (black), -0.136 V/nm (green) and -0.27 V/nm (blue). The solid grey lines are a fit to the curve, and the dashed horizontal line is a guide the eye to show where the fitted curve cuts the y-axis. (f) Charge inhomogeneity ($\delta n$) calculated from the intersection of the fitted curves to the minimum conductivity as a function of $D$. The shaded region with data points in blue shows similar data for the control device of graphene on monolayer WSe$_2$.}
\label{transport2}
\end{figure*} 
\section{Methods}
 2D layers were obtained by mechanical exfoliation technique, identified through optical contrast under a microscope and verified by Raman spectroscopy. Two monolayers of WSe$_2$ with sharp edges making an of 60$^\circ$, corresponding to zigzag edges \cite{debnath2021simple, guo2016distinctive}, were chosen to fabricate the twisted bilayers. We used dry transfer method to pick-up and transfer the monolayers, with one layer effectively rotated near $\sim2.1^\circ$ with respect to the other. The details of the edge alignment during the dry transfer process are discussed in our previous work \cite{debnath2021simple} and in SI, section I. The electrical leads are fabricated after patterned etching using reactive ion plasma followed by Cr/Au (5nm/50nm) metallization. Fig.~\ref{transport1}(c(i)) shows the cross-sectional view of the device geometry which is in a hall-bar configuration where the effective channel is on monolayer graphene with a top-gate architecture and a global back-gate on SiO$_2$ (285 nm)/Si$^{++}$ substrate. $L = 4 \mu$m and $W = 3\mu$m are the length and width of the hall-bar channel. Fig.~\ref{transport1}(c(ii)) shows the optical micrograph of the fabricated device. The heterostructure consists of graphene/twisted WSe$_2$ (Gr/tWSe$_2$) stack fully encapsulated by hexagonal boron nitride (hBN) crystals of thicknesses $\sim20$ nm and $\sim40$ nm at the bottom and top respectively. The presence of dual gates enable us to tune the number density ($n$) and the displacement field ($D$) independently. $n$=[$C_{tg}(V_{tg}-V_{t0})+C_{bg}(V_{bg}-V_{b0})$]$/e$ and    $D(V/nm)=[C_{tg}(V_{tg}-V_{t0})-C_{bg}(V_{bg}-V_{b0})]/2\epsilon_0$ where, $C_{bg}$ and $C_{tg}$ are bottom and top gate geometric capacitance respectively, $V_{tg}$ and $V_{bg}$ are the applied top and bottom gate voltages and $V_{b/t0}$ is the gate voltage corresponding to the intrinsic doping. Another similar device with monolayer WSe$_2$ was prepared as a control experiment by the same fabrication process, which is discussed in SI, section III. 

\section{Results}
\subsection{1. Electrical characterization of Gr/tWSe$_2$ FET}
Fig.~\ref{transport1}(d) shows the transfer characteristics of the graphene FET with $V_{bg}$ and $V_{tg}$. The $V_{bg}$ sweep shows a clear and prominent hysteresis at room temperature, which is absent during the $V_{tg}$ sweep. The hysteresis in the $V_{bg}$ sweep is reproducible, present after multiple thermal cycles, and remains unchanged for $V_{bg}$ sweep rates ranging from 100 V/hr to 900 V/hr. Hysteresis due to the charge trapping centers in substrates or surface adsorbates are ruled out because (a) our device was fully encapsulated with hBN and exhibited reasonably high channel mobility ($\sim 26,000$~cm$^2$/V-s), (b) top gate sweep shows no hysteresis, and (c) presence of local charge traps generally leads to an {\it anti-}hysteresis behavior \cite{feng2014temperature,wang2010hysteresis,joshi2010intrinsic}. On the other hand, the transport data in  Fig.~\ref{transport1}(d)(i) are similar to previous reports of interfacial ferroelectricity on graphene-TMDCs heterostructures \cite{weston2022interfacial,wang2022interfacial,yasuda2021stacking}. (Possibility of ferroelectric domains in similarly prepared devices is supported from piezo-responsive force microscopy given in SI, section II.) Since the twist angle ($\theta \approx 2.1^\circ$) lies close to the {\it anti-}ferroelectric-dielectric crossover, the observation of the hysteresis itself possibly indicates contribution from the regions of lower twist angles due to unavoidable angle inhomogeneity. The estimated polarization ($P_{2D} \approx$ 0.18$\times$10$^{-12}$ C/m) from the peak separation is accordingly a factor of $\sim 5 - 10$ smaller than typical values of $P_{2D}$ in twisted TMDCs \cite{weston2022interfacial, zhang2023ferroelectric, li2017binary}, which could also be due to polarization switching occurring only partially in the anti-ferroelectric regime (due to presence of a large percentage of domain walls of topological nature that prevents a complete transformation of all the domains \cite{ko2023operando}).
   
\subsection{2. Effect of displacement field on Gr/tWSe$_2$}
 We examine next the combined influence of the bottom and top gates on the channel properties, specifically investigating the effects of an out-of-plane electric field. We continuously sweep $V_{tg}$ and hold $V_{bg}$ at a fixed voltage (see Fig. \ref{transport2}(a)) and vice versa at 80 K (i.e. where there were no hysteresis between the forward and reverse sweeps). In the $R-V_{tg}$ sweeps we make two unique observations: (i) the resistance of Dirac point reduces with increasing $V_{bg}$, and (ii) there is a slight broadening of the $R-V_{tg}$ curve at large $V_{bg}$ values. Fig.~\ref{transport2}(b) shows $V_{bg}$ sweeps at fixed $V_{tg}$ which reveals additional complexity in the form of multiple Dirac points that become more prominent with increasing magnitude of $D$, as we go to larger $V_{tg}$. At lower temperatures, the anomalous resistance peaks become more pronounced (shown in SI, section IV). Similar measurements performed on the control device with monolayer WSe$_2$ placed below graphene does not show any of the above anomalies (see SI, section III). Fig.~\ref{transport2}(c) and (d) show the color plot of $R$ in the $D-n$ and $V_{tg}-V_{bg}$ spaces, respectively, where multiple lines, each corresponding to a local resistance maximum, fan out from the origin signifying an emergent disorder, where the channel disintegrates into regions of different charge densities, through introduction of the external electric field. This is distinct from intrinsic disorder, {\it e.g.} from inhomogeneous charge distribution due to Coulomb impurities, where the evolution of local Dirac points would form parallel lines in the $V_{tg}-V_{bg}$ phase space. Interestingly, while the main locus (denoted as D1 in Fig.~2(d)) is determined by purely geometric gate-channel capacitances with the slope $= C_{tg}/C_{bg} \approx d_{b}/d_{t}$ ($d_{b}$ and $d_{t}$ are measured dielectric thickness between the graphene channel and the backgate ((hBN/SiO$_2$) and topgate (hBN), respectively), the other minor lines (D2, D3 and D4) seem to indicate a stronger back-gate coupling, {\it i.e.} enhanced $C_{bg}$ upto a factor of $\sim 10$ (D4).

 To quantify the extent of emergent charge inhomogeneity ($\delta n$) we extrapolate the conductivity ($\sigma$), normalized by its minimum value ($\sigma_{min}$), to evaluate the density at which channel conductivity becomes independent of $n$ \cite{nam2017electron, nagashio2013estimation}. $\delta n$ then corresponds to the intersection of the extrapolated lines onto the $n$-axis at $\sigma/\sigma_{min} = 1$ (red dashed line in Fig.~2(e). In Fig.~\ref{transport2}(f) we plot $\delta n$ as a function of $D$, which clearly establishes strong enhancement in charge inhomogeneity at large electric fields. With inhomogeneity in the graphene/monolayer WSe$_2$ control heterostructure being expectedly low ($\sim$1$\times10^{11}$/cm$^2$) and independent of $D$ (blue markers in Fig.~2(f)), the striking emergent inhomogeneity can be directly associated to the presence of the twisted TMDC bilayer in the back-gate stack. 
\begin{figure*} 
\includegraphics[width=1\linewidth]{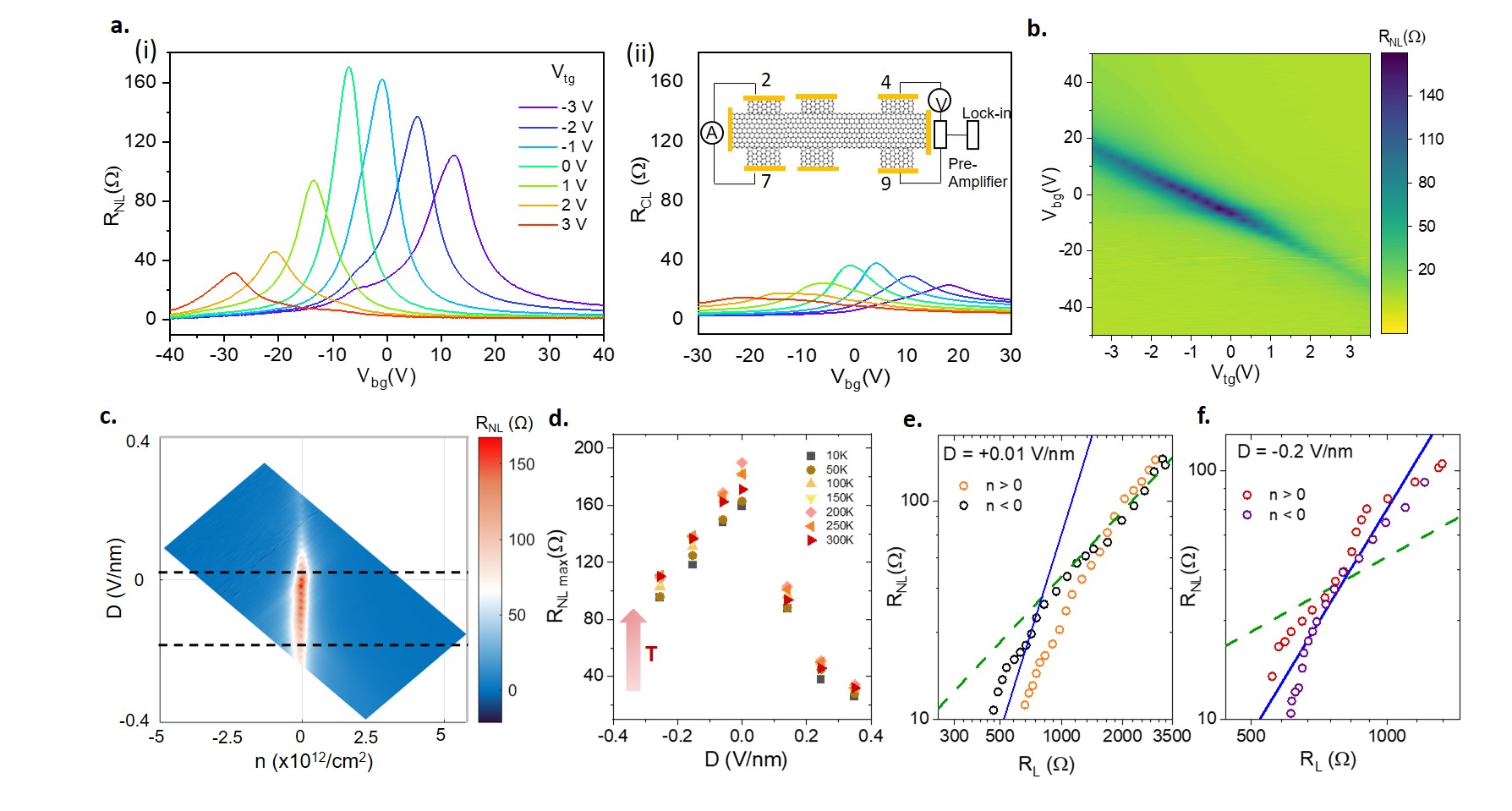} 
\caption{ (a) Non-local measurements in Gr/tWSe$_2$. (i) Non-local resistance of graphene and (ii) the classical contribution to the non-local resistance as a function of back-gate voltage at different top gate voltages at room temperature. Inset shows the contact configuration and the measurement setup. (b) The color plot of $R_{NL}$ as a function of V$_{tg}$ and V$_{bg}$. (c) Color map of $R_{NL}$ as a function of total carrier density $n$ and the displacement field $D$. The horizontal y-cuts indicate the fixed values of $D$ for which the scaling analysis was performed. (d) Temperature dependence of $R_{NL}$ at Dirac point with $D$ from 300 K to 10 K. (e,f)  $R_{NL}$ as a function of $R_{L}$ at fixed $D$ values of -0.01 V/nm and 0.2 V/nm respectively. The solid blue line and the dashed green line represents cubic and linear scalings, respectively.}
\label{transport3}
\end{figure*}

\begin{figure}[thb]
\centering  
 \includegraphics[width=\columnwidth]{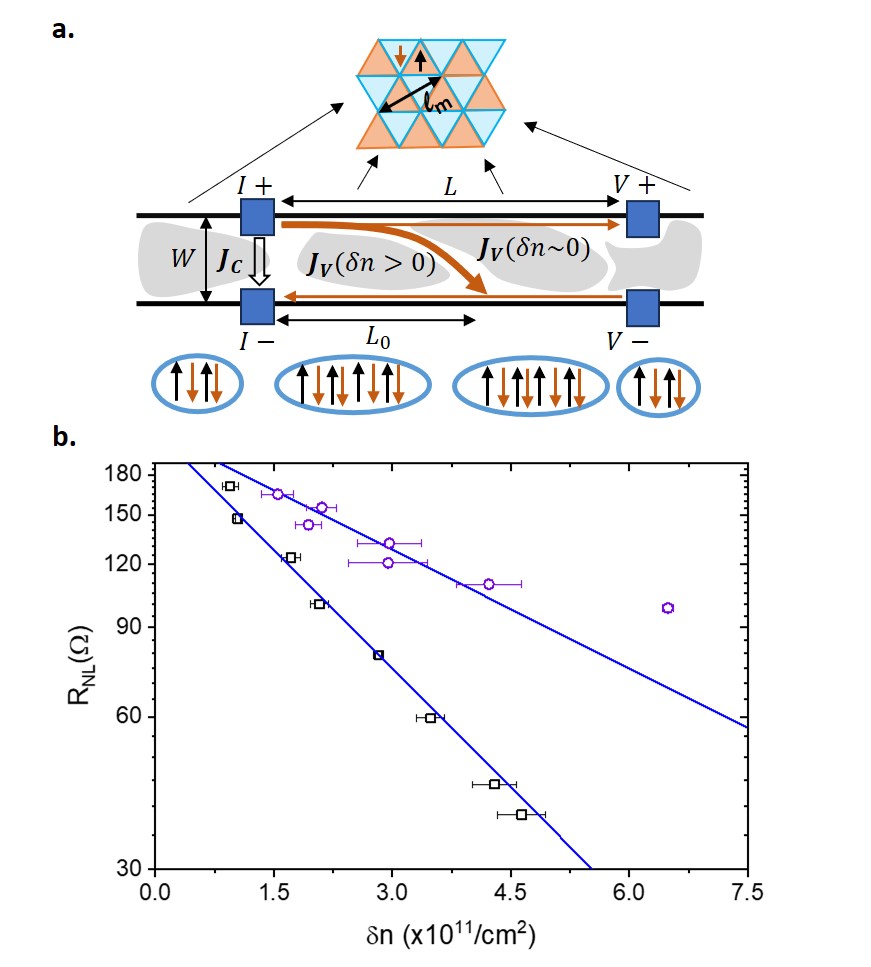} 
\caption{(a)  Schematic of the phenomenological model, correlating the valley current $J_V$ with inhomogeneity $\delta n$. The usual path of charge transport is denoted by $J_C$ that flows between the current contacts, $I_+$ and $I_-$. In the presence of a displacement field, some of the ferroelectric domains expand, that disrupts the triangular domain pattern, while increasing the charge inhomogeneity fluctuations experienced by graphene.(b) $R_{NL}$ as a function of inhomogeneity ($\delta n$) for $D$ $>$ 0 V/nm (black square) and $D$ $<$ 0 V/nm (purple circle). The solid blue line is fit to the data.}
\label{transport4}
\end{figure}
\subsection{3. Non-locality in the graphene channel}
In order to obtain a deeper understanding of the emergent background potential fluctuations, we carried out detailed non-local (NL) transport measurements in our device. The measurement layout is schematically shown in the inset of Fig.~\ref{transport3}(a). We choose two probes (4,9) furthest away from the current-injecting contacts (2,7) to measure the non-local voltage so as to minimize the classical Ohmic contribution to resistance (See SI, section IV, for experimental detail). A pronounced non-local resistance ($R_{NL}$) is observed in Gr/tWSe$_2$ (shown in Fig.~\ref{transport3}(a)) at all temperatures. In Fig.~\ref{transport3}(a.ii), we show the expected classical Ohmic resistance contribution $R_{CL} = (R_{L}W/L)\exp(-\pi L/W)$, where $W$ and $L$ are the width of the Hall bar and the distance between the current and voltage leads, respectively. The non-local resistance can be  as much as $\sim$ten times larger than what is expected from the classical Ohmic contribution. No non-local signal was observed on control graphene/monolayer WSe$_2$ devices with similar device geometry (SI, section III). 

To explore the dependence of $R_{NL}$ on the gate electric fields, we show the color plot of R$_{NL}$ (at $T = 12$~K) in the $V_{tg}-V_{bg}$ space in Fig.~\ref{transport3}(b). Interestingly, the satellite peaks observed for the longitudinal resistance $R_L$ (Fig.~2(d)) are absent in the $R_{NL}$ map, implying that $R_{NL}$ is immune to the bulk disorder in the channel. For a more quantitative analysis, we recast $R_{NL}$ in the $D-n$ space in Fig.~\ref{transport3}(c). The graph clearly shows that $R_{NL}$ peaks near the CNP and at $D \approx 0$ with a maximum value of $\approx$ 160 $\Omega$ while it decreases rapidly for both positive and negative $D$. However, the overall behaviour is slightly asymmetric as the rate of decay in $R_{NL}$ with $D$ depends on the direction of $D$ irrespective of temperatures (Fig. \ref{transport3}(d)), which points to the possibility that the non-locality is connected to the nature and extent of the electrostatic coupling between the tWSe$_2$ layers.

\section{Discussions}
The opposite trends in the emergent charge inhomogeneity and the non-local resistance with varying electric field suggest both effects could after all be manifestations of the same underlying microscopic mechanism. The non-locality is usually associated with unconventional physical phenomena such as edge modes, topological spin and valley currents, hydrodynamic flows etc. and associated with a symmetry breaking mechanism \cite{abanin2011giant,balakrishnan2013colossal,volkl2019absence,wei2016strong,stepanov2018long,ribeiro2017scale,wei2018electrical,nachawaty2018large,tanaka2021charge,sui2015gate,shimazaki2015generation, gopinadhan2015extremely, hung2020experimental,wu2019intrinsic, sinha2020bulk,ma2020moire, gorbachev2014detecting}. In  graphene-based devices, such as bilayer graphene subjected to a transverse displacement field \cite{sui2015gate,shimazaki2015generation} or monolayer graphene rotationally aligned on the hBN substrate \cite{gorbachev2014detecting}, two competing mechanisms involve topological valley Hall currents driven by bulk Berry curvature due to the broken inversion symmetry \cite{xiao2007valley} or quasi-dispersive chiral edge modes in gapped graphene that is not necessarily topological in nature \cite{kirczenow2015valley, marmolejo2018deciphering, aktor2021valley}. Often the scaling of $R_{NL} \propto R_{L}^\alpha$ provides insight into the mechanism of non-local transport, where $\alpha = 3$ for pure valley/spin current \cite{sui2015gate, shimazaki2015generation, gorbachev2014detecting, abanin2009nonlocal}, while on the other hand, the edge mode transport yields a linear scaling, {\it i.e., }$\alpha = 1$, where the non-local signal decays exponentially as $R_{NL} \sim R_L\exp[-\pi L/L_0]$, where $L_0$ is a characteristic relaxation length scale which is much longer than the geometric width of the Hall bar itself~\cite{srivastav2024electric, brown2018edge}. 

In Fig.~\ref{transport3}(e) and (f) we plot $R_{NL}$ with $R_{L}$ for two fixed values of $D$; 0.01 V/nm and -0.2 V/nm, respectively, where the magnitudes of $R_L$ were extracted from Fig.~\ref{transport2}(c) corresponding to value of $n$ along the dashed lines in Fig.~\ref{transport3}(c). For $D \approx 0$~V/nm (Fig.~\ref{transport3}(e)), the scaling is evidently linear, although it approaches a cubic scaling for lower $R_L$ ({\it i.e.} far away from the charge neutrality). For large values of $D$, the magnitude of $R_{NL}$ is considerably lower and the scaling was found to be largely cubic ($\alpha \simeq 3$). Although the cubic scaling relation could be indicative of pure valley hall effect with globally broken A/B sub-lattice symmetry \cite{sui2015gate, gorbachev2014detecting} - a possibility that we do not rule out because of the presence of the WSe$_2$ layer in van der Waals proximity to the graphene channel, it could also be mediated by the edge current~\cite{marmolejo2018deciphering}. While the pure valley current in metal dichalcogenide heterostructures can propagate several tens of micrometers \cite{jin2018imaging}, the relaxation in edge-bound valley current in graphene, for example, in strongly gapped bilayer graphene \cite{aamir2018marginally}, occurs at a much shorter length scale ($\sim 0.5~\mu$m) due to defects in edge termination morphology, and thus cannot contribute to the non-local transport significantly (width of our Hall bar, $W \approx 3~\mu$m). Thus the edge mode transport in our device, inferred from the dominance of the linearly scaling component of $R_{NL}$ at small $D$, requires a different source of inversion symmetry breaking that is relatively more immune to the boundary disorder. We thus conclude that the tessellated network of polarized domains in the twisted WSe$_2$ bilayer that can lead to varying regions of broken sublattice symmetry in the graphene channel, and thereby a quasi-dispersive edge-bound valley current, is the most likely origin of the non-local resistance in our device.

We now discuss the apparent opposite behaviour of the inhomogeneity $\delta n$ and non-locality $R_{NL}$ as a function of $D$ within the framework of the electric field-dependent ferroelectric texture of the underlying tWSe$_2$ layer. At $D \sim 0$, the ferroelectric polarization oscillates at the moir\'{e} scale ($\ell_m \approx 9$~nm for $\theta = 2.1^\circ$ in our device) over macroscopic regions of the sample allowing the edge-bound valley current to propagate over a long distance. With increasing $D$, larger regions of polarized domains disrupt such short wavelength fluctuations in the polarization. The unavoidable spatial inhomogeneity in the electrostatic coupling between tWSe$_2$ and graphene layers, assisted by the variation in the twist angle and disorder-mediated domain wall pinning etc., makes the channel increasingly inhomogeneous, and causes $\delta n$ to increase with $D$ (as observed in Fig.~\ref{transport2}(f)). Simultaneously, this will force the coherently dispersing edge modes to deviate from the sample boundary and follow a preferred domain morphology (Fig.~\ref{transport4}(a)), thereby reducing the valley relaxation length ($L_0$) and suppressing $R_{NL}$. Since the valley polarization will invariably be lost at the opposite physical boundaries of the Hall bar, we estimate $L_0/W \simeq N_m \approx 1/\ell_m^2\delta n$, where $N_m$ is the  number of moir\'{e} cells that forms a coherent domain along which the edge mode can propagate, and thus represents a characteristic scale for the valley current propagation. When $N_m$ is large ($\delta n \approx$ 0), the valley current mostly propagates along the physical edge (large $L_0$), whereas  at large inhomogeneity ($\delta n \approx$ 10$^{15}$/m$^2$), $N_m$ is smaller causing deflection in the valley current. This causes $R_{NL}$ to decrease exponentially with $\delta n$ as $R_{NL} \propto \exp[-\beta\delta n]$ with $\beta = \pi\ell_m^2 L/W \approx 0.34\times10^{-15}$m$^2$. Fig.~\ref{transport4}(b) shows the variation of $R_{NL}$ with $\delta n$ extracted from their values at the same $D$ (from Fig.~\ref{transport2}(f) and Fig.~\ref{transport3}(d)). In addition to the evident exponential decay, the solid lines in Fig.~\ref{transport4}(b) indicate $\beta  = 0.36\times10^{-15}$m$^2$ and $0.18\times10^{-15}$m$^2$ for the positive and negative values of $D$, respectively, which are close to that expected from the phenomenological model.

Finally, we wish to comment on the observation of multiple subsidiary Dirac points in Fig.~\ref{transport2}(d) that evolve with capacitive couplings stronger than the geometric capacitance between the back-gate and the graphene channel. Capacitive amplification with a ferroelectric gate stack is a well-discussed topic \cite{salahuddin2008use, khan2015negative}. Recent theoretical and experimental studies show that a multi-domain system can support stable negative capacitance, leading to capacitive amplification \cite{hoffmann2018stabilization, yadav2019spatially, zubko2016negative}. Alternately polarized moir\'{e} domains in twisted WSe$_2$ naturally offers a multi-domain system with a similar free-energy landscape \cite{bennett2023polar} in which stable negative capacitance may exist leading to enhanced capacitance. It is not surprising that the magnitude of capacitance amplification is not uniform throughout, because it will depend on multiple factors including the angle inhomogeneity, which vary strongly across the different regions of the device.

\section{Conclusion}
We investigate electrical transport in a field-effect architecture in graphene placed on a near-parallel bilayer of twisted bilayer WSe$_2$. The device exhibits multiple unexpected features that include emergent charge inhomogeneity, anomalous resistance features and temperature insensitive non-local transport. We show quantitatively these effects are inter-related and likely to be manifestations of the anti-ferroelectric moir\'{e} domains in the underlying twisted bilayer. Additionally, the back-gate-channel capacitance appears to exhibit local enhancement which could also be another facet of the (anti)-ferroelectricity incorporated in the (back) gate stack. Our experiment uncovers several unique aspects of ferroelectric proximity-driven electrical transport in graphene via valley current manipulation. 

\section{Contributions}
S.S. and R.D fabricated the device, performed the measurements, analysed the data and wrote the manuscript. A.S. contributed partially to measurements and analysis of data. S.M. and M.J. performed the theoretical calculations. J.K.M., V.R., M.B. and G.S. performed sample characterizations. K.W. and T.T. grew the hBN crystals. A.G. supervised the overall project and edited and revised the manuscript.

\section{ACKNOWLEDGMENTS}
The authors acknowledge the usage of NNFC facilities at CeNSE, IISc and funding from Department of Science and Technology (DST), Govt. of India. A.G. acknowledges the J. C. Bose Fellowship (Grant number SP/DSTO-18-2038) from Science and Engineering Research Board, DST, Govt. of India. K.W. and T.T. acknowledge support from the JSPS KAKENHI (Grant Numbers 20H00354 and 23H02052) and World Premier International Research Center Initiative (WPI), MEXT, Japan.

\section{Supporting Information}
\section{I. Methods}
\subsection{A. Transfer process for orientationally accurate twisted heterostructures}
2D layers were mechanically exfoliated and verified through Raman Spectroscopy. Fig. 1(a) shows the two monolayer TMDCs with zig-zag edges identified and marked. The angle between two sharp edges of the flake is usually 60$^\circ$. We align the sharp edges and twist them to a near-parallel configuration. The two monolayers after the transfer process is shown in Fig. 1 (b). The twist angle from the optical image is approximately 5$^\circ$. 

We used second-harmonic generation (SHG) technique to determine the twist angle between the two monolayers after the transfer process. Polarization-dependent SHG measurements were performed using a nonlinear microscopy imaging setup. The system utilized a linearly polarized fundamental excitation at 1040 nm from a mode-locked fiber laser (Coherent Fidelity HP-10 laser, 80 MHz repetition rate, and pulse duration of 140 fs) as an excitation source. The input laser was scanned using a galvo-scanner (GVS001, Thorlabs) and was focused onto the sample using a 20 $\times$0.75 NA objective lens, and the same objective was used to collect the backward emitted SHG signal. The SHG signal was detected using a photomultiplier tube (PMT, Hamamatsu R3896) with a dichroic filter, a set of bandpass (520/15 nm) and short pass (890 nm) filters, and a polarizer (analyzer) mounted in front of it for rejecting the fundamental source and minimizing the background noise. For the twist angle measurement, the sample was rotated from 0–90$^\circ$ in steps of 5$^\circ$ with respect to the laboratory horizontal axis, keeping the input polarizer and the output analyzer fixed and perpendicular to the laboratory horizontal axis. For each angle, SHG images of the twisted sample were acquired and later analyzed to calculate the twist angle.

Fig. 1(c) shows the second-harmonic generation (SHG) mapping shows the homobilayers, with maximum intensity of the near-parallel twisted region while the monolayers show a lower intensity. The net SHG emission from the stacking region will be a coherent superposition of layer intensities along with a phase difference dependent on the stacking angle. Constructive or destructive interference occurs between the two SHG emission depending on the stacking angle of 0$^\circ$ and 60$^\circ$ respectively. The intensity polar plot of the monolayers shows a twist angle of $\approx$ 5$^\circ$. Dots are the experimental data points and the solid lines are the curve fit for the two monolayers. Fig. 1(d) show the room temperature Raman spectra of twisted WSe$_2$ with the characteristic A$_{1g}$/E$_{2g}$ peak at 241 cm$^{-1}$ and a small peak (B$_{1g}$) corresponding to the breathing mode at 308 cm$^{-1}$, which is absent in the monolayer. Recent studies have shown that the ratio of $B_{1g}/A_{1g}$ can be used to approximately determine the twist angle. A ratio $\approx$ 0.01-0.05 indicates twist angle $<$ 5$^\circ$ as is true in this case as well. Using this transfer technique and by careful observation of the sharp zigzag edges, we can optically determine the twist angles of the homobilayers.

In a similar process, we proceed to fabricate the heterostructure with the layers arranged as follows:hBN-graphene-WSe$_2$-WSe$_2$-hBN on SiO$_2$/Si substrate. Fig. 2(a) shows an optical image of the transferred hetero-stack with multiple layers. A color corrected image helps us identify the individual layers as marked by an outline. The zigzag edges used to align the two monolayers are marked as well. The twist angle determined from here is 2.1$^\circ$. 
\begin{figure}[h!]
\includegraphics[width=1\linewidth]{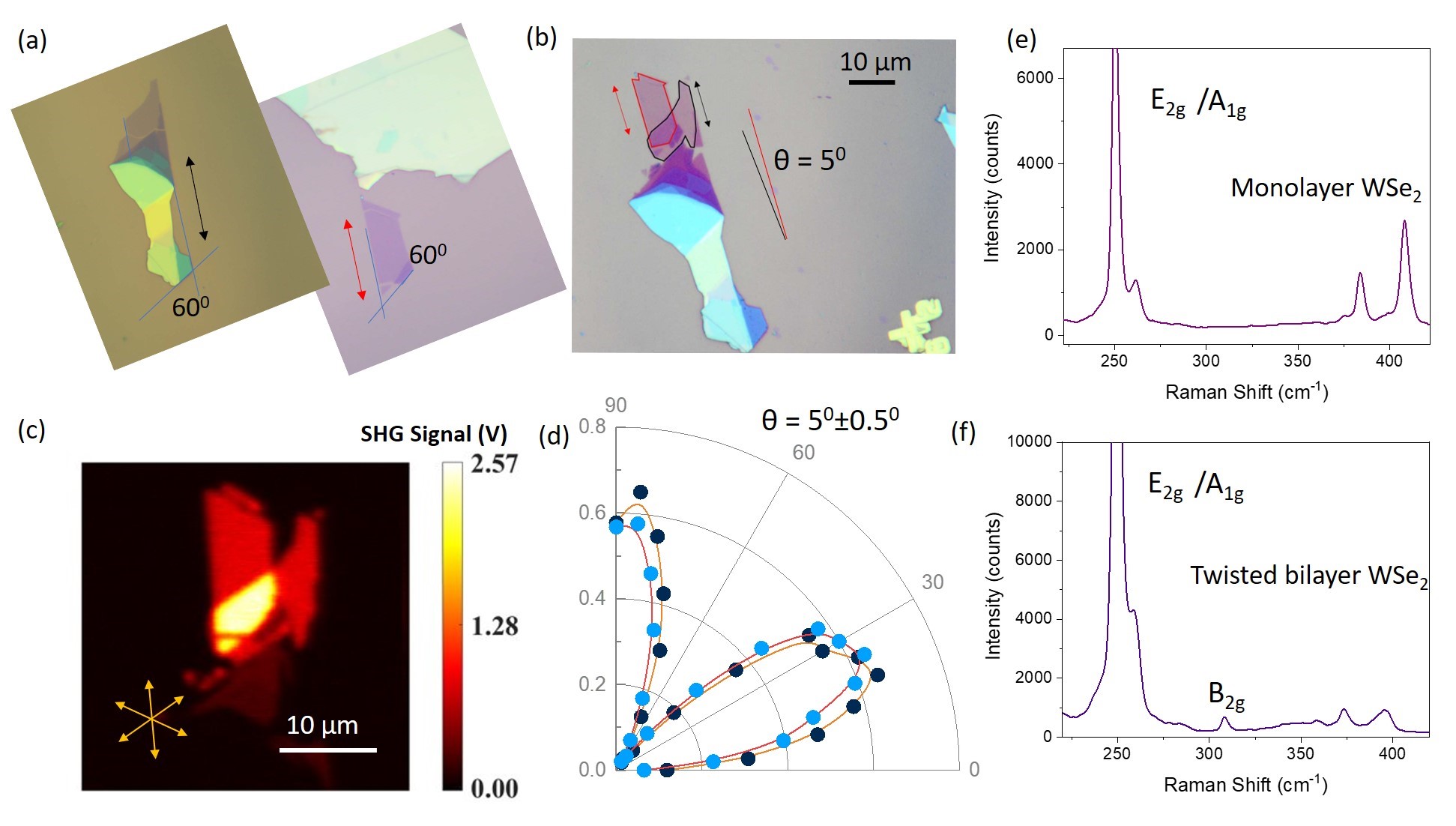} 
\caption{Transfer process for near-parallel rotational alignment.(a) Optical image of monolayers with zigzag edges marked with arrow pointers. (b) Optical image of the artificial heterostructure after the transfer. The zig zag edges of the individual monolayers are at a twist angle of $\approx$ 5$^\circ$. (c) The SHG map of twisted bilayer WSe$_2$, where the x and y axis are the spatial directions and the color represents the SHG voltage signal. The yellow arrows indicate the zig-zag direction. (d) SHG polar plot of the monolayers in the heterostructure. The experimental data points are indicated by circles. The top flake is dark blue and the bottom flake is light blue. Fits to the experimental data (red curve) shows a twist angle of 5$^\circ$ in between the two flakes. (e) Raman Spectroscopy of monolayer WSe$_2$ and (f) twisted bilayer WSe$_2$\ showing the relevant modes.}
\label{S1}
\end{figure}
\begin{figure}[h!]
\includegraphics[width=0.9\linewidth]{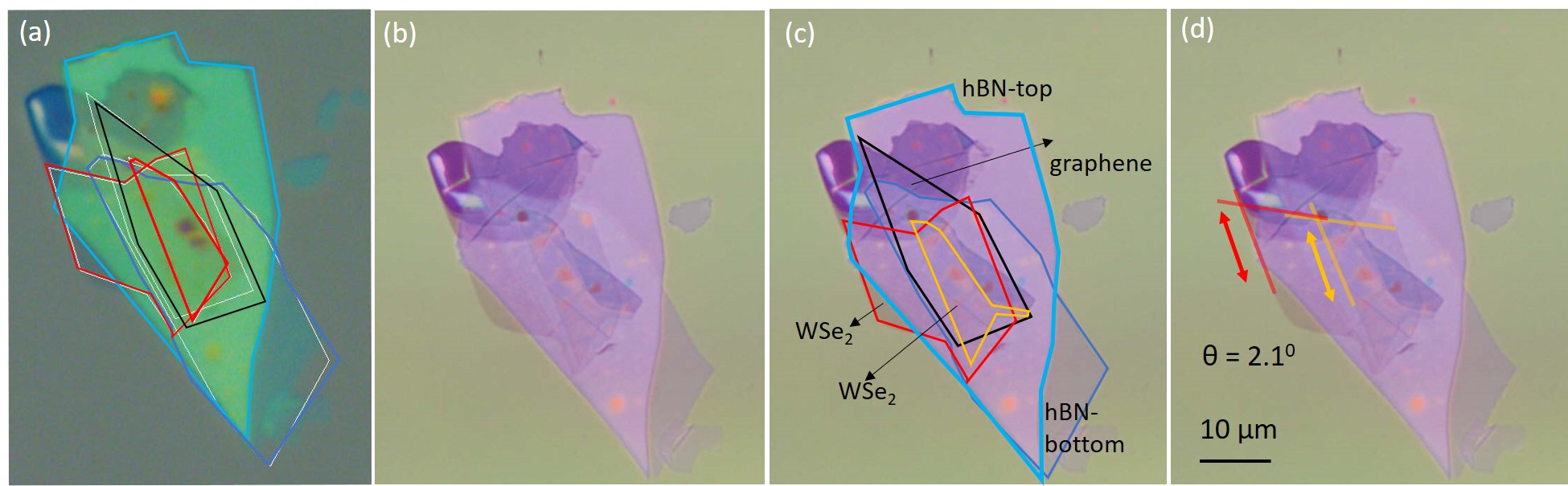} 
\caption{Optical images of the van der Waals heterostructure.}(a) During dry transfer process to fabricate the heterostructure with each layer outlined. (b) Color corrected image of the same in which the individual layers can be visually distinguished.(c) Individual flakes outlined and labelled in the heterostructure. (d) The zigzag edges marked with arrow pointers which were used for rotational alignment.
\label{S2}
\end{figure}
\begin{figure}[h!]
\includegraphics[width=0.85\linewidth]{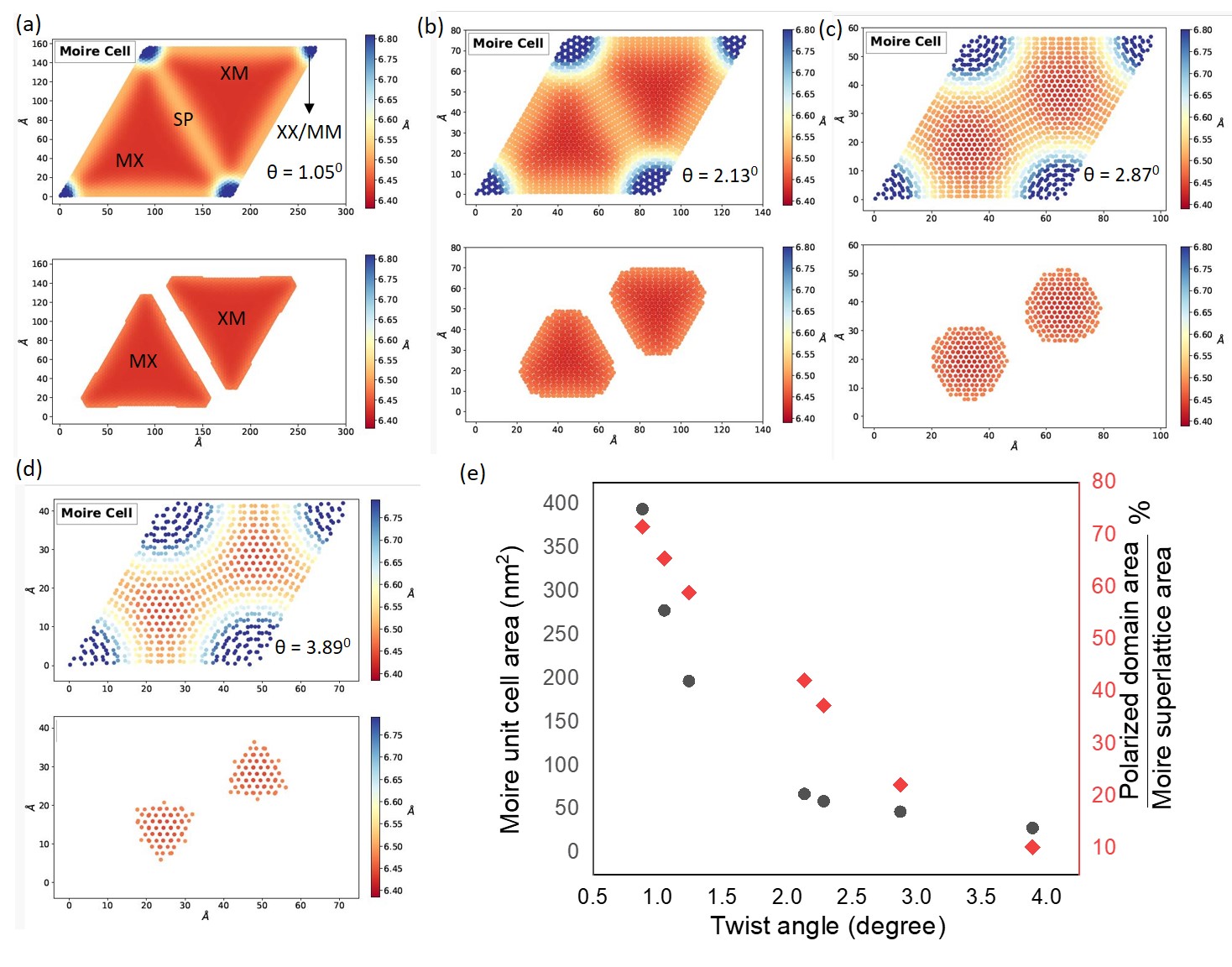} 
\caption{(a-d) Color plots of interlayer separation in a moire cell with twist angle showing how the stacking area evolves with twist for near-parallel alignment in an artificially stacked bilayer WSe$_2$. (e) The dependence of moire cell size and stacking area of MX/XM domains.}
\label{dft}
\end{figure}
\subsection{B. Basic Electrical Setups}
 We used an a.c. current at $\approx$ 223 Hz of 100 nA for 4-point probe electrical measurements using a lock-in amplifier. The encapsulated device with non-invasive device architecture has a field effect mobility ($\mu$) of $\sim 26000$ cm$^2$/Vs and $\sim 15000$ cm$^2$/Vs at hole and electron side respectively at 80 K indicating a good quality transport channel. For non-local measurements we use a slightly different scheme. Measurement artifacts may appear when typical lock-in measurement setup is used for detection of non-local signal. The spurious signal may come from the common-mode voltage at the current injection side of the Hall bar. Since the lock-in amplifier has an input impedance of $\approx$ 10M$\Omega$ (SR830 Lock-in amplifier), the common-mode voltage could induce a current through the two non-local leads. To eliminate this issue, we used a pre-amplifier SR551 that has an input impedance of more than teraOhm at a low frequency of around 17 Hz, the output of which went into the lock-in amplifier.

\section{II. Structural properties of domains in twisted WSe$_2$}
\subsection{A. First principles calculations on commensurate moire superlattice in near-parallel twisted bilayer WSe$_2$}
The rigid structures for twisted bilayer WSe$_2$ were generated using the TWISTER \cite{naik2022twister} code. The systems were relaxed in LAMMPS \cite{thompson2022lammps} using Stillinger Weber \cite{mobaraki2018validation} as the intralayer potential and Kolmogorov Crespi \cite{naik2019kolmogorov} as the interlayer potential upto a force tolerance of $10^{-6}$ $eV/\AA$. The different stacking regions were isolated in the relaxed structures by investigating the interlayer separation. Fig. 3(a-d) shows the different stacking regions in near-parallel twisted bilayer WSe$_2$ for twist angles of 1.05$^\circ$, 2.13$^\circ$, 2.89$^\circ$ and 3.89$^\circ$. The color code indicates the interlayer separation (ILS) between the two monolayers, which is a result of repulsion between atomic orbitals dependent on the atomic registry and leads to lattice reconstruction/relaxation. Distinct solitonic region formations were observed in the relaxed structures. The MM (metal on metal) stacking regions have the highest interlayer distance while the MX/XM (metal on chalcogen and chalcogen on metal) regions have the lowest. We have isolated the MM and the MX/XM stacking regions as the regions having interlayer distance greater than 6.7$\AA$ and less than 6.45$\AA$ respectively. Then remaining parts of the moir´e unit cells are denoted as the saddle point or the SP regions. There is a significant modification in the lattice reconstruction from 1.05$^\circ$ to 3.89$^\circ$. In Fig. 3(e) we plot the moire unit cell area and the percentage stacking area of polarized moir\'{e} domains as a function of low twist angles. Twisted bilayers show negligible lattice reconstruction beyond 4$^\circ$.

It is to be noted that at a twist angle $\theta\approx2.1^\circ$, the domain wall width between MX and XM domains is significant due to large area fraction of these domain walls as compared to perfect parallel stacking. The topologically protected domain wall network (DWN) governs the expansion of a particular domain while extrinsic factors like bubbles, defects etc prevent domain wall motion causing pinning of a domain. At twist angles between $\theta_{c}$ and $\theta_{t}$, the moir\'{e} superlattice does not become completely ferroelectric, rather remains anti-ferroelectric with presence of a small fraction of oppositely polarized domain even at a sufficiently large fields \cite{ko2023operando}. The DWN in twisted bilayers are topologically protected as multiple DWs meet at the MM regions or nodes, forming a perfect partial dislocation and hence cannot be eliminated \cite{ko2023operando, weston2022interfacial}. These combined factors result in an inhomogeneous medium of MX and XM domains at large $D$.

\subsection{B. Piezoelectric Force Microscopy on twisted bilayer WSe$_2$}
Fig. 4(a) shows a twisted bilayer TMDC device fabricated specifically for Piezoelectric Force Microscopy (PFM) measurements. The arrangement of individual layers are given alongside. The layers are arranged such that the bottom hBN lies in on a graphite flake which is in contact with micro-sized gold electrodes. We mounted the sample on a metallic plate with conducting silver paint. The sample was grounded properly to avoid charging effects. To perform PFM experiments, the required area of the sample was cleaned mechanically by an AFM tip. The cantilever with Pt/Ir coating and a force constant of  $\sim$ 0.7- 9.0 N/m was used. The cantilever was tuned in non-contact mode at its resonance frequency. The non-contact resonance frequency of the cantilever was found to be 77 KHz. The topography of the sample was first traced in AC non-contact mode. The same area was traced in contact mode slowly by decreasing the set point (0 V) and deflection (-0.4). A conducting tip was used to scan the sample surface. The cantilever was oscillated by applying AC bias to the piezo block attached to the cantilever. Further, the cantilever was tuned at the contact resonance frequency. All the measurements were performed at this resonance frequency to get maximum sensitivity. PFM measures the mechanical piezo-response to the applied bias between the tip and the sample. The sample locally expands and contracts according to the applied tip bias, which changes the deflection of the cantilever. The piezo-response signal of the sample was recorded in terms of the first harmonic component of the deflection of the cantilever. For the measurements, a sinusoidal AC voltage was applied to the piezo attached to the cantilever chip to drive the cantilever in oscillation. The cantilever was  then tuned at its contact resonance frequency which  was found to be 340 KHz at which all the measurements were done.

Fig. 4(b-c) shows successive magnified topographic images of twisted bilayer WSe$_2$. Finally, the phase map in Fig.4(d) is collected at a satisfactory magnification enabling us to visualize the alternate triangularly polarized domains which have been outlined in red. There is a dark portion followed by a bright portion alternately arranged. These patches are the polarized moire regions, separated by regions which are unpolarized. The PFM being sensitive to strain, and the other region shows a usual moire pattern formation with criss-cross structures.
\begin{figure}[h!]
\includegraphics[width=0.7\linewidth]{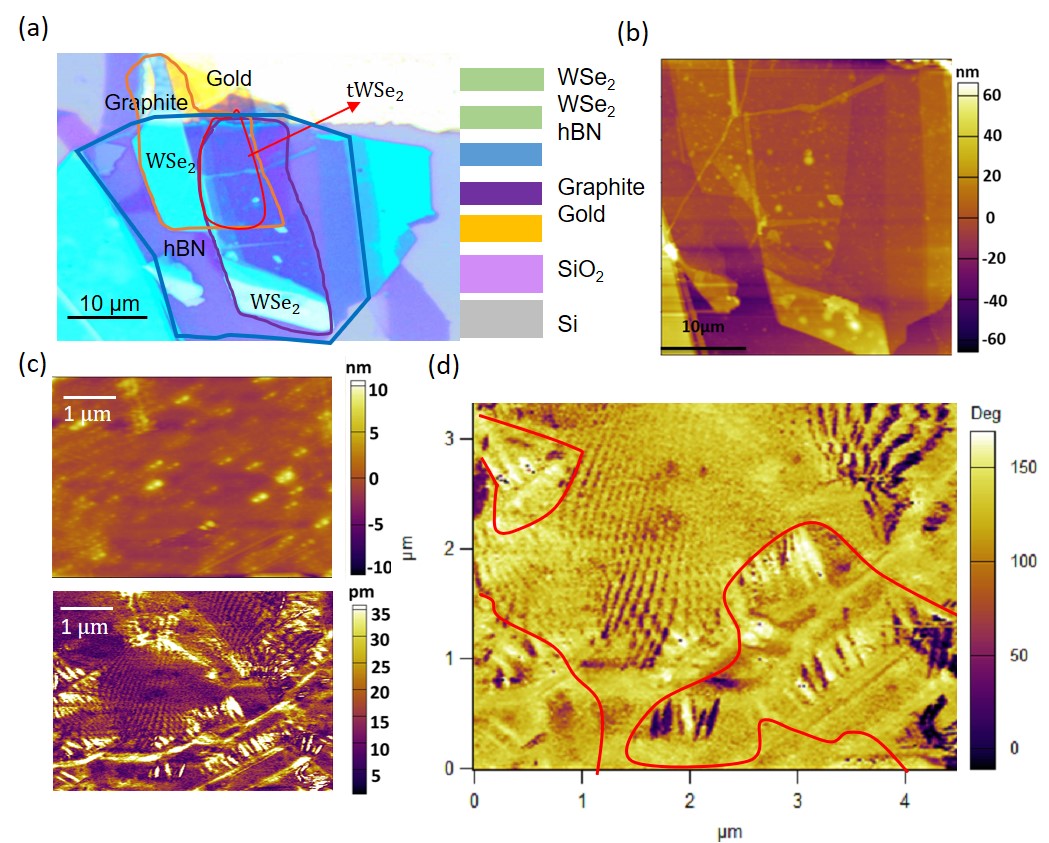} 
\caption{PFM measurements on twisted bilayer WSe$_2$ at room temperature. (a)Heterostructure for PFM measurements and arrangement of 2D layers given alongside. (b) AFM topography map of the heterostructure where the different flakes can be visualized by height contrast. (c) AFM and PFM image of the twisted WSe$_2$ region. (d) Magnified view of the PFM phase plot showing bright and dark contrast of polarized domains.}
\label{S3}
\end{figure}
\section{III. Measurements on Gr/WSe$_2$ monolayer control device}
We fabricated an hBN encapsulated graphene with monolayer WSe$_2$ inserted between graphene and hBN. The heterostructure was etched in a hall bar geometry with dual gate architecture in a field-effect configuration as shown in Fig. 5(a). We performed similar electrical measurements in this device. Fig. 5(b) shows the forward and backward traces in Gr/WSe$_2$. There is no hysteresis present. Fig. 5(c-d) shows gate sweeps with incremental V$_{tg}$ and V$_{bg}$. It shows no anomalous behaviour in the resistance with monolayer WSe$_2$/hBN as the back-gate stack. The color plots in Fig.5(e-f) show that there is only one Dirac point in both gate sweeps. From these measurements we can claim that monolayer WSe$_2$ has no impact on the graphene channel with respect to charge doping or capacitive amplification.

Non-local measurements in the control device were performed in a similar manner as explained in Section I. Fig. 5(g) show the usual $R_{L}$ as a function of back-gate voltage. In Fig. 5(h) we plot $R_{NL}$ and $R_{NL}$-$R_{CL}$, to show that the effective non-local signal is negligible. Hence, there is no emergent non-locality in the control device of graphene on monolayer WSe$_2$.
\begin{figure}{h!}
\includegraphics[width=1\linewidth]{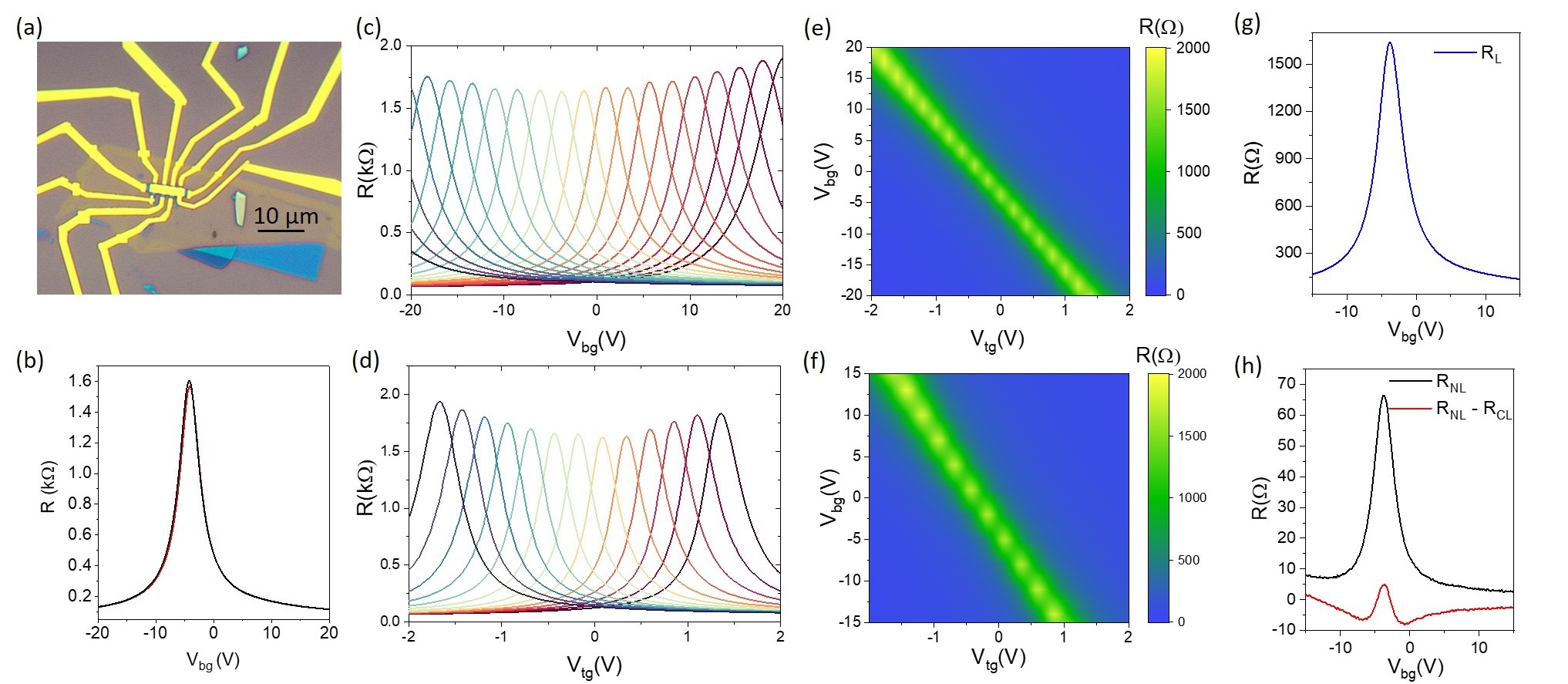} 
\caption{Characterization of Gr-WSe$_2$ control device. (a) Optical image of the dual-gated device in a hall bar geometry. (b) R-V$_{bg}$ curves for forward and backward traces at 300 K. (c) V$_{bg}$ sweeps for incremental V$_{tg}$ and (d) V$_{tg}$ sweeps for incremental V$_{bg}$. (e-f) Color plots for V$_g$ sweeps at 80 K. (g) $R_L$ for V$_{bg}$ sweep (h) $R_{NL}$ (black) and $R_{NL}-R_{CL}$ (red), calculated using the graph given in (g).}
\label{S4}
\end{figure}

\section{IV. Extended electrical measurements on Gr/tWSe$_2$}
\subsection{A. Measurements related to hysteresis and inhomogeneity}
\begin{figure}{h!}
\includegraphics[width=1\linewidth]{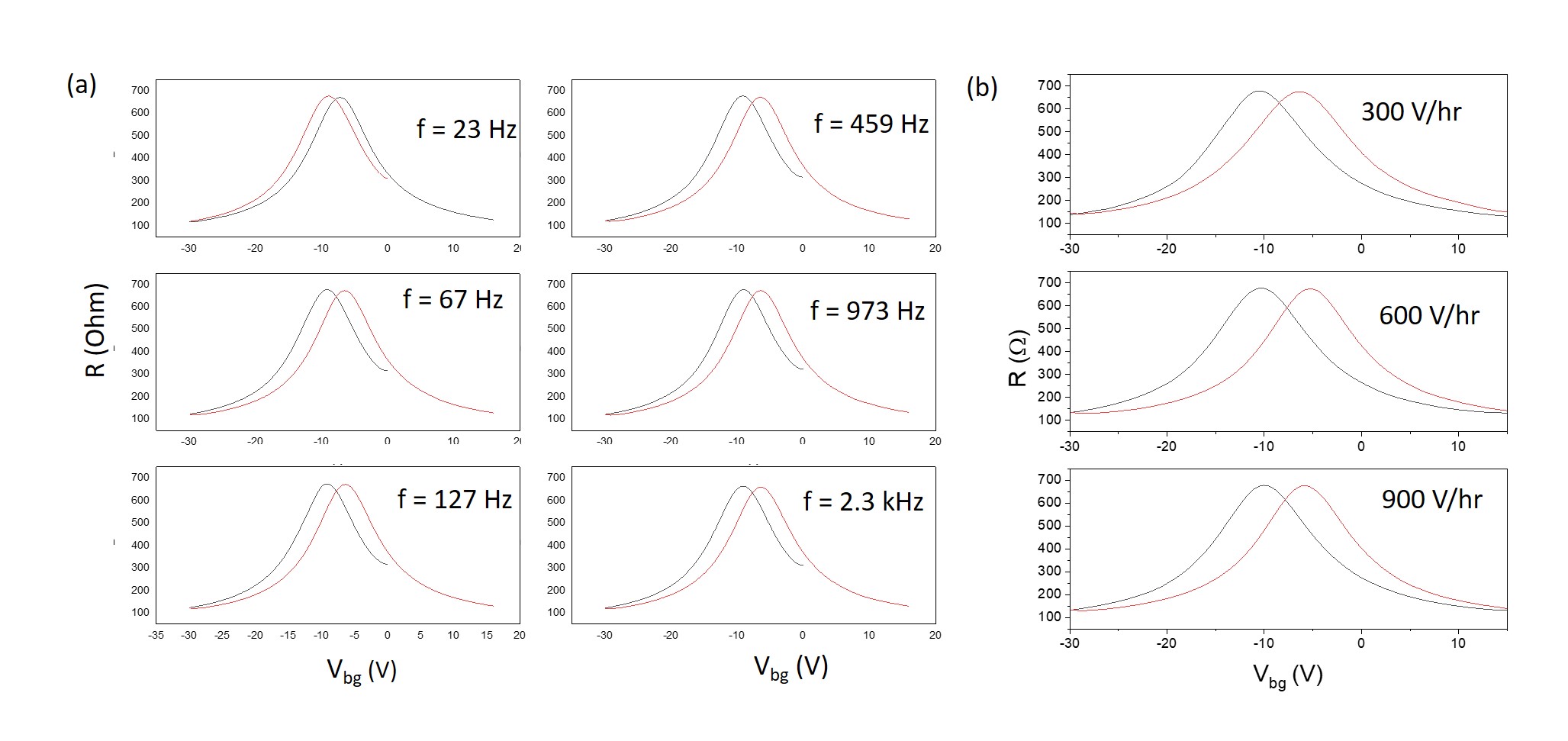} 
\caption{Hysteresis in Gr-tWSe$_2$(a) $R-V_{bg}$ curves at 300 K for input frequencies from 23 Hz to 2.3 kHz. (b) Variation of sweep rate from 300 V/hr to 900V/hr.}
\label{S5}
\end{figure}
Fig. 6(a) shows R-V$_{bg}$ curves at 300 K for different input frequencies.The hysteresis remains over two orders of magnitude in the a.c. input frequency. The hysteresis also remains with V$_{bg}$ sweep rate increasing from 300 V/hr and 900 V/hr. There is no reduction in the induced charge, or change of hysteresis direction. We can conclude that indeed ferroelectricity induced charge doing leads to hysteresis. 
\begin{figure}{h!}
\includegraphics[width=0.7\linewidth]{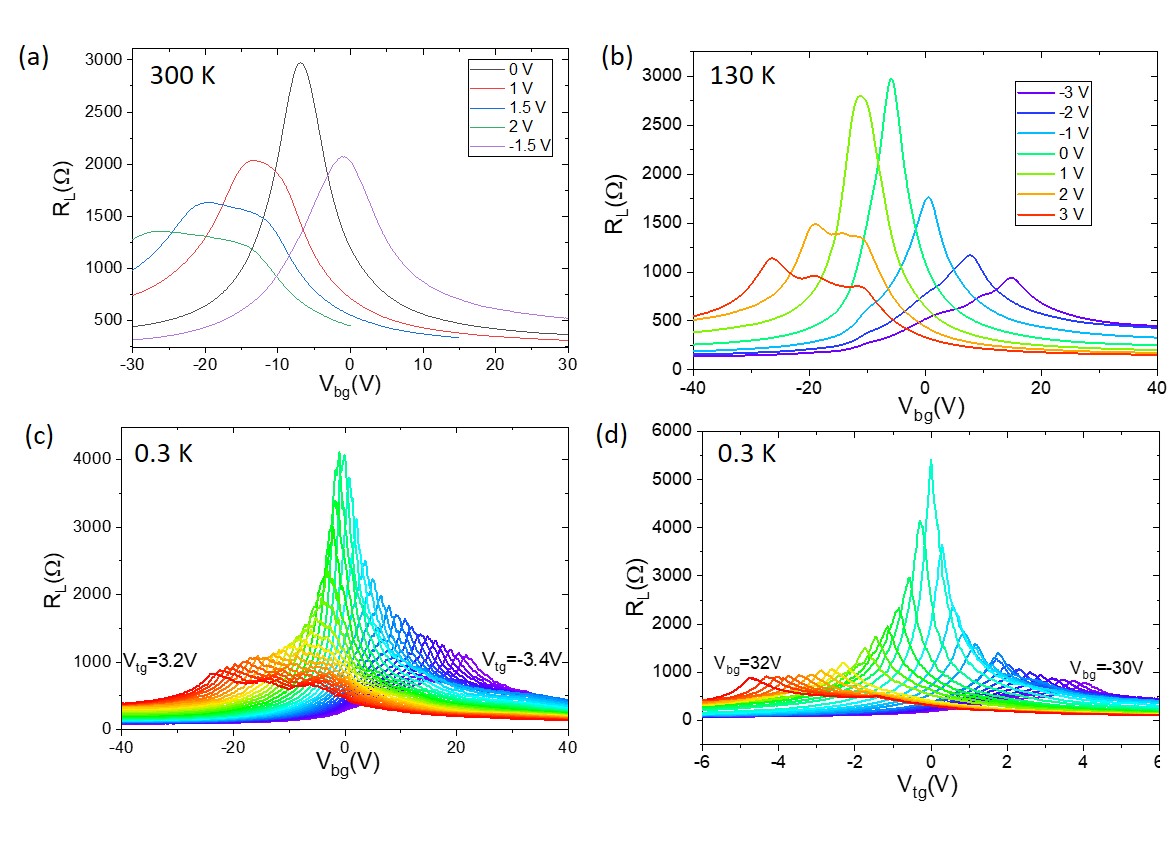} 
\caption{$V_{bg}$ sweeps for different $V_{tg}$ values at (a) 300 K, (b) 130 K and (c) 0.3 K. V$_{tg}$ sweep at 0.3 K.}
\label{S6}
\end{figure}
There is charge inhomogeneity throughout the explored temperature range of 0.3K to 300K in Gr/tWSe$_2$ (see Fig. 7(a-c)). It is more prominent at low temperatures. The inhomogeneity is partially screened by graphene during the top gate sweep as shown in Fig.7(d). This is another indication charge inhomogeneity arises from the underlying ferroelectric. It is temperature independent right up to 300 K which shows the robustness of the twisted structure.

\subsection{B. Additional non-local measurements in Gr/tWSe$_2$}
\begin{figure}{h!}
\includegraphics[width=1\linewidth]{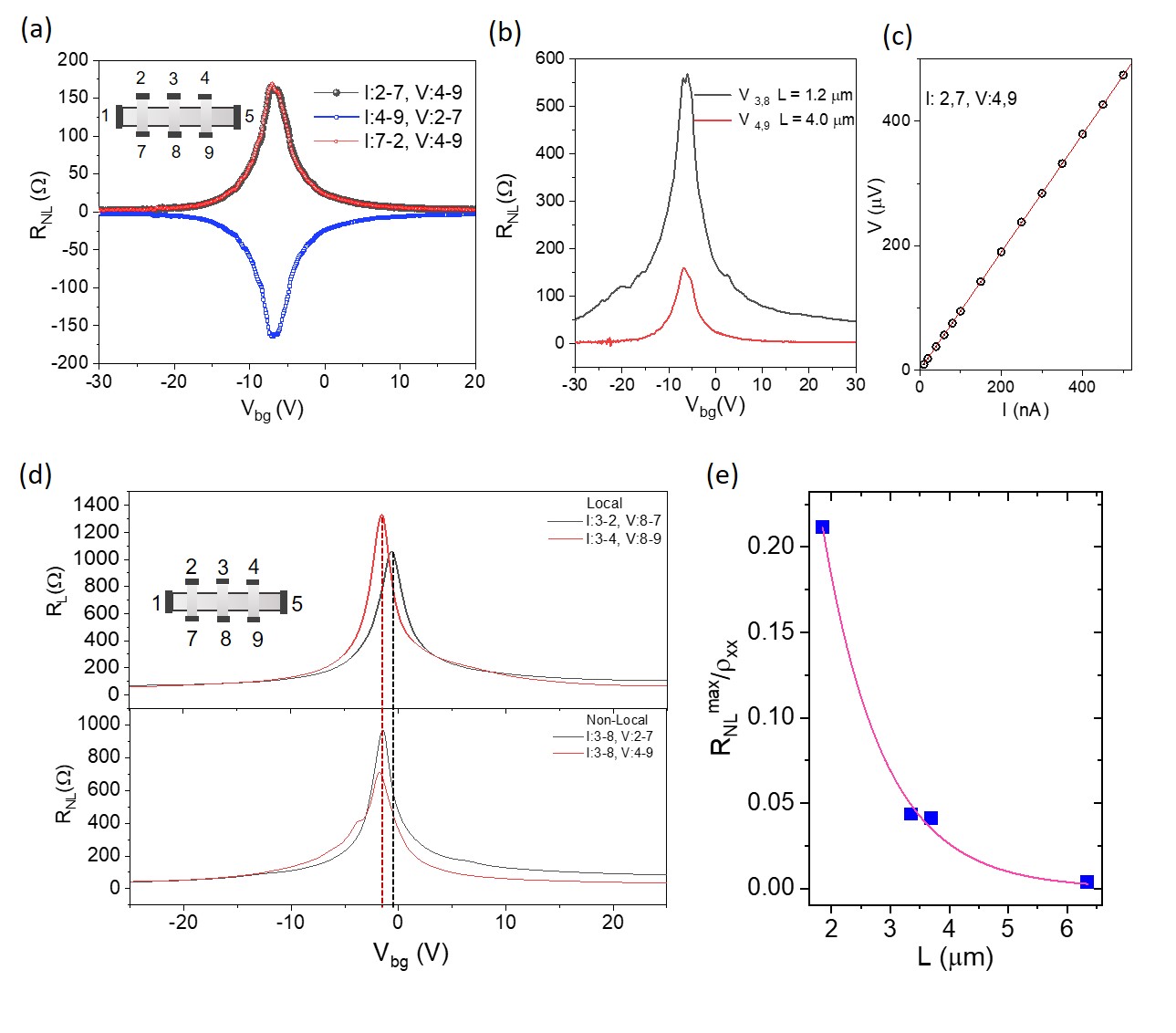} 
\caption{(a) $R_{NL}$ in different channels and with current and voltage probes interchanged and reversed. The measurement configurations are mentioned in the legend. (b) Non-local signal for different contact lengths. (c) Non-local voltage as a function of current, showing a linear response. (d)  $R_L$ and $R_{NL}$ taken for different contact configurations as shown in the inset. The dashed line marks the peak position for $R_L$, through to the lower panel. (e) The maximum non-local resistance as a function of channel length in a different Gr/tWSe$_2$ sample. The solid line is an exponential fit to the curve.}
\label{S7}
\end{figure}
\begin{figure}{h!}
\includegraphics[width=0.9\linewidth]{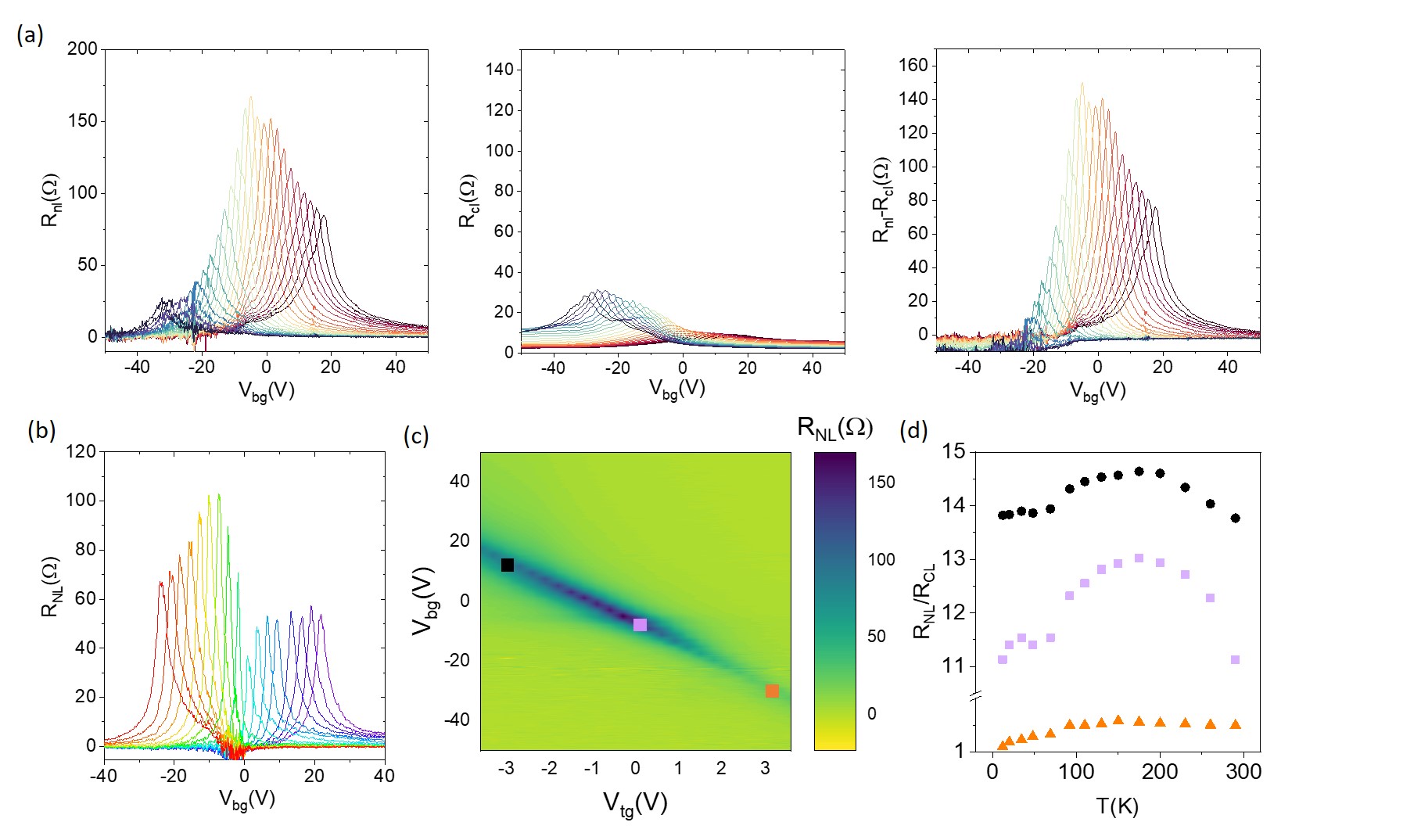} 
\caption{(a) Non-local measurements in Gr/tWSe$_2$ at 12 K. The three panels showing $R_{NL}, R_{CL}, R_{NL}-R_{CL}$. (b) Non-local signal $R_{NL}$ at 300 mK. (c) Phase space plot of the same as a function of V$_{tg}$ and V$_{bg}$. (d) Non-local signal at three values of $D$ (0.14 V/nm, 0 V/nm and -0.14V/nm) as a function of temperature.}
\label{S8}
\end{figure}
\begin{figure}[h!]
\includegraphics[width=0.4\linewidth]{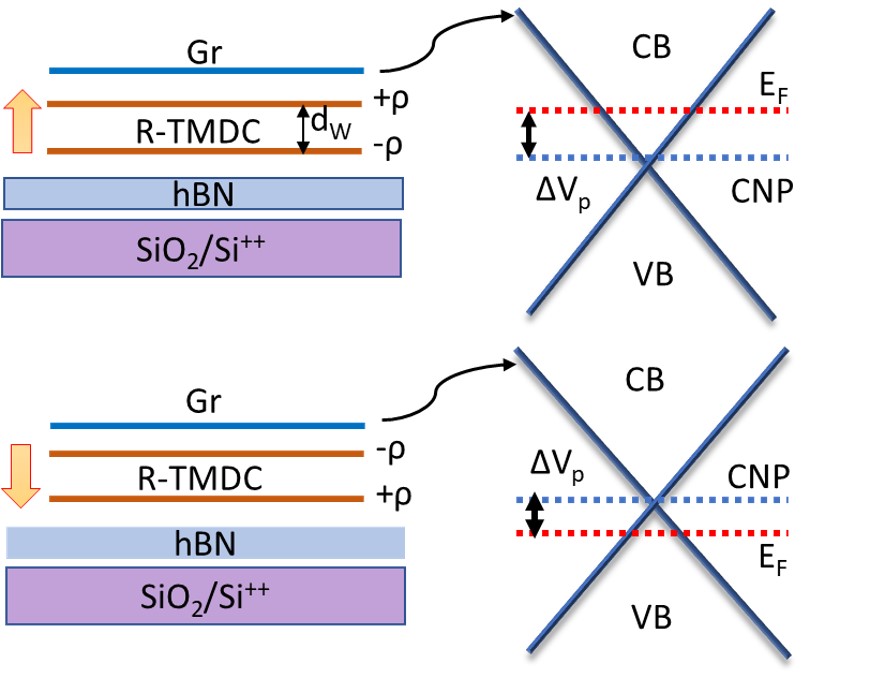} 
\caption{Schematic explaining the nature of hysteresis in Gr/tWSe$2$ due to opposite polarization effects on application of a displacement field. The built-in voltage is $V_p$ in graphene due to charge $\rho$ accumulation in between the ferroelectric layers of WSe$_2$ with gap $d_W$ = 0.65 nm.}
\label{S9}
\end{figure}
We performed a set of basic checks on the non-local voltage measurements to rule out artifacts potentially from other sources such as heating effects, common-mode voltage, etc. We observed the non-local voltage changed its sign when the local current injection was reversed indicating that the heating effect was negligible in our setup. In addition, our non-local measurement satisfied the universal reciprocal relation, i.e. we observed the same signal if the current and voltage probes were swapped. Moreover, to make sure that the observed non-local signal is not due to any thermo-electric effect like a contribution from Joule heating, we have studied the dependence of the non-local voltage as a function of the applied current. If the dominant contribution is from the thermo-electric effect, the voltage should show a non-linear dependence with the current.
In Fig. 8(a), shows three non-local voltage measurements for different current and voltage probe configurations: usual non-local configuration, with reversed current and voltage probes, and with voltage and current leads swapped. The non-local signal is consistent. In Fig 8(b) we study $R_{NL}$ as a function of length in between the current injecting probes and the voltage probes. With increasing length, the non-local signal reduces since the classical contribution varies with distance between voltage and current leads, due to diffusive transport, given by: 
\begin{equation}
    R_{NL}=  \frac{4}{\pi} \rho_{xx} exp(\frac{-L}{W}) 
\end{equation}
as we observe from the fit of exponential decay given in Fig. 8 (e).
In Fig.(c) we plot $R_{NL}$ as a function of excitation current from 10 nA to 500 nA. The non-local response is linear, indicating no heating effects or thermo-emf \cite{sui2015gate, shimazaki2015generation}. Moreover, the temperature gradient due to the thermo-electric effect should be along the length of the sample, while the measured non-local voltage is across the width in the Hall bar geometry, which also allows us to exclude any such contribution from the thermo-electric effect on the measured non-local signal.Thus, we ascertain that the observed non-locality is an emergent feature of graphene twisted WSe$_2$. We next determine if there are any adverse effects to the non-local signal due to sample inhomogeneity. Sample inhomogeneity manifests as shift of resistance peaks at CNP. The broken lines mark the peak positions of $R_L$ (Fig. 8(d)) for two different contact configurations. The different peak position in $R_L$ and $R_{NL}$ obtained from different parts of the sample is a result of sample inhomogeneity \cite{blake2009influence}. There is no other effect of it on the non-local signal. Having performed all necessary non-local measurements we rule out possibilities of spurious non-local signals and ascertain the validity of non-locality in our system.

Additional non-local measurements were performed in Gr/tWSe$_2$. Fig. S9(a) shows non-local($R_{NL}$), classical ($R_{CL}$) and $R_{NL}-R_{CL}$ resistance in Gr/tWSe$_2$ at 12 K. Fig. 9(b) shows $R_{NL}$ at 0.3 K. The color plot in Fig. 9(c) at 12 K shows only one Dirac point. At three displacement field values, marked by squares in the color plot, we determine the non-local resistance. We plot them as a function of temperature in the form of $R_{NL}/R_{CL}$ in Fig. S9(d). 

\section{V. Polarization, charge inhomogeneity and Capacitance calculations}
\subsection{A. Calculation of polarization from shift of CNP}

The difference in back-gate voltage corresponding to the CNP for forward and backward sweep ($\Delta V$) is necessary to compensate for polarization induced charge that gives rise to hysteresis. A finite electric field exists between TMDC and graphene due to the built-in interlayer potential in twisted bilayer WSe$_2$ even at zero field. Thus, an extra bottom gate voltage of $\Delta V$ is required to align the Fermi level of graphene ($E_F$) to the charge neutrality point (CNP) when the polarization (P) is up (down) as shown schematically in Fig. 10. We would like to clarify that our device architecture is different from the previous device structures; here, the ferroelectric layer lies on top of $\approx$ 20 nm thick hBN and a 285 nm SiO$_2$ dielectric. Hence, the charge induced is due to the entire ferroelectric/dielectric gate stack of twisted WSe$_2$/hBN/SiO$_2$ (thickness: $d_{total}$ $\approx$ 300 nm). So, induced charge is modified as: 
\begin{equation}
   2\Delta n = \epsilon\varepsilon \frac{\Delta V}{d_{total}}
\end{equation}
We get, $\Delta n$ = 3.53x$10^{11}$/cm$^2$. We derive the expression of the 2D polarization $P_{2D}$ that can be modeled as two layers of positive and negative charge densities($\rho\pm$) separated by distance between the two WSe$_2$ monolayers, $d_{W}$, hence $P_{2D} =\pm \rho d_{W}$. Since $\rho = \Delta n e$, we get,
 \begin{equation}
     P_{2D}= \Delta n e d_{W} 
 \end{equation}
and a corresponding, 2D polarization $P_{2D}$ = 0.18x$10^{-12} C/m$ for twisted WSe$_2$ at $\sim2.1^\circ$ rotation.

We provide a table below with important device parameters from previous reports in graphene on interfacial 2D ferroelectrics and compare it with our work.
\begin{table*}[ht]
\caption{Device design and ferroelectric parameters for twisted 2D bilayers.}
\begin{tabular}{|p{15mm}|p{15mm}|p{17mm}|p{15mm}|p{15mm}|p{17mm}|} 
 \hline
 Report& Material &Architecture &Total Induced charge 2$\Delta$ n (cm$^2$) &Electric field due to shift in CNP$\Delta/d$ (V/nm) &Polarization P$_{2D}$ (C/m) \\
 \hline
Yasuda et. al (expt.) \cite{yasuda2021stacking} &0$^\circ$ twisted bilayer hBN &hBN back-gate &2x10$^{11}$ &- &0.64x10$^{-12}$ \\
\hline
Zhang et. al (expt.) \cite{zhang2023visualizing} &0$^\circ$ twisted bilayer WSe$_2$ &SiO$_2$ (285 nm) global back-gate &7.5x10$^{11}$ &0.04 &0.78x10$^{-12}$\\
\hline
Wang et.al (expt.) \cite{wang2022interfacial} &0$^\circ$ twisted bilayer WSe$_2$ &hBN back-gate &- &0.018 &2x10$^{-12}$\\
\hline
Weston et.al (expt.) \cite{weston2022interfacial} &0$^\circ$ twisted bilayer WSe$_2$ &hBN back-gate &1.25x1012 &- &0.6x10$^{-12}$\\
\hline
Li et. al (theory) \cite{li2017binary} &0$^\circ$ twisted bilayer MoS$2$ &- &9x10$^{11}$ &- &0.97x10$^{-12}$\\
\hline
This work (expt) &2.1$^\circ$ twisted bilayer WSe$_2$ &hBN (20 nm) + SiO$_2$ (285 nm) global back-gate &3.53x10$^{11}$ &0.017 &0.18x10$^{-12}$\\
\hline
\end{tabular}

\end{table*}

We would like comment on the polarization magnitude obtained by us, since transport is not the ideal technique to measure the polarization magnitude. It is also true that polarization switching does not occur throughout the device as has been demonstrated by Ko et. al. \cite{ko2023operando}. This is indeed consistent with our results since we estimate $P_{2D}$ to be a factor of five to ten lower than what is expected theoretically.
\begin{itemize}
    \item The value of polarization amplitude calculated from the CNP shift is just an estimate and it does not affect/alter any of the primary claims of the paper, namely, the emergent electric field-induced inhomogeneity in graphene and the observation of non-locality in transport. 
    \item None of the above were observed in our control experiments with monolayer WSe$_2$ in the bottom gate stack (instead of the twisted bilayers).
    \item The experimentally observed magnitude of the polarization is only about 10 \% of the theoretically expected magnitude at zero-degree twisted bilayers due to spatially inhomogeneous ferroelectric switching and to twist angle of 2.1 $^\circ$, that is close to the ferroelectric-dielectric crossover. 
\end{itemize}
 

\subsection{B. Calculation of charge inhomogeneity}
To determine the ferroelectricity induced charge doping in graphene, we need to understand the process by which polarized moire domains influence graphene. The opposite doping in graphene occurs in spatially separated domains, which causes the Fermi energy to oscillate in space resulting in electron-hole ‘puddles’. This is the root cause of the inhomogeneity that we observe in the experiment. The oppositely polarized domains cause a fluctuation in the fermi level of graphene, mimicking a disordered potential landscape, giving rise to charge inhomogeneity and this effect does not cancel out. In the presence of usual disorder, e.g. due to Coulomb impurities, there are equally probable regions of electron-rich and hole-rich puddles at the charge neutrality \cite{martin2008observation}. Thus the spatially separated doping of opposing sign do not cancel out, but increases inhomogeneity, and determines the minimum conductivity ($\sigma_{min}$) in graphene. Hence, using double-logarithmic plot of the conductance ($\sigma$) as a function of carrier density ($n$), we extract the charge inhomogeneity ($\delta n$). We extrapolate the conductivity from the regime where $\sigma \propto n^k$ and determine the charge inhomogeneity where this line intersects $\sigma_{min}$. Some representative plots at different values of displacement field $D$ for this device is given in Fig. 2(d) of main manuscript. In Fig. 11 we show the analysis for our control device at three different $D$ values of 0, -0.19 and +0.255 V/nm. For graphene on monolayer WSe$_2$, we observe an almost constant value of $\delta n$, ranging from 0.85x$10^{11}$-1.4x$10^{11}$/cm$^2$. 

We would like to comment on the extent of ferroelectric doping: we could expect it to be $\approx$ 10$^{11}$/cm$^2$ in our system, which determines the scale of inhomogeneity that we observed in the experiments. Typical value of $P_{2D}$ in twisted TMDCs is $\approx$ 0.6-2x10$^{-12}$C/m (references in table above) which corresponds to an induced ferroelectric doping of $\approx$ 1.2-4x10$^{12}$/cm$^2$ when it is placed on top of graphene (d = 0.3 nm, distance between the graphene and the ferroelectric layer). In our case, the strength of $P_{2D}$ extracted from the hysteresis in transport is significantly weaker because:
\begin{itemize}
    \item We are close to anti-ferroelectric to dielectric crossover.
    \item Angle inhomogeneity occurs because of the transfer process in twisted van der Waals hybrids.
    \item Polarization switching occurs only partially in the antiferroelectric regime, due to various reasons, ranging from spatial inhomogeneity of twist angle, to  topological nature of the domain walls that prevents a complete transformation of all the domains.
\end{itemize}
The theoretical value of $P_{2D}$ = 0.97x10$^{-12}$C/m  in twisted MoS$_2$ \cite{li2017binary}. From here, we can approximately estimate the average polarization switching area to be $\approx$ 20 \%, which is not a disagreeable estimate considering a finite global twist angle of 2.1$^\circ$ that is at a near critical value. As a result, the expected charge inhomogeneity that is induced by the ferroelectric layer should be $\approx$ 10$^{11}$/cm$^2$ as has indeed been observed experimentally.

\begin{figure}[h!]
\includegraphics[width=0.7\linewidth]{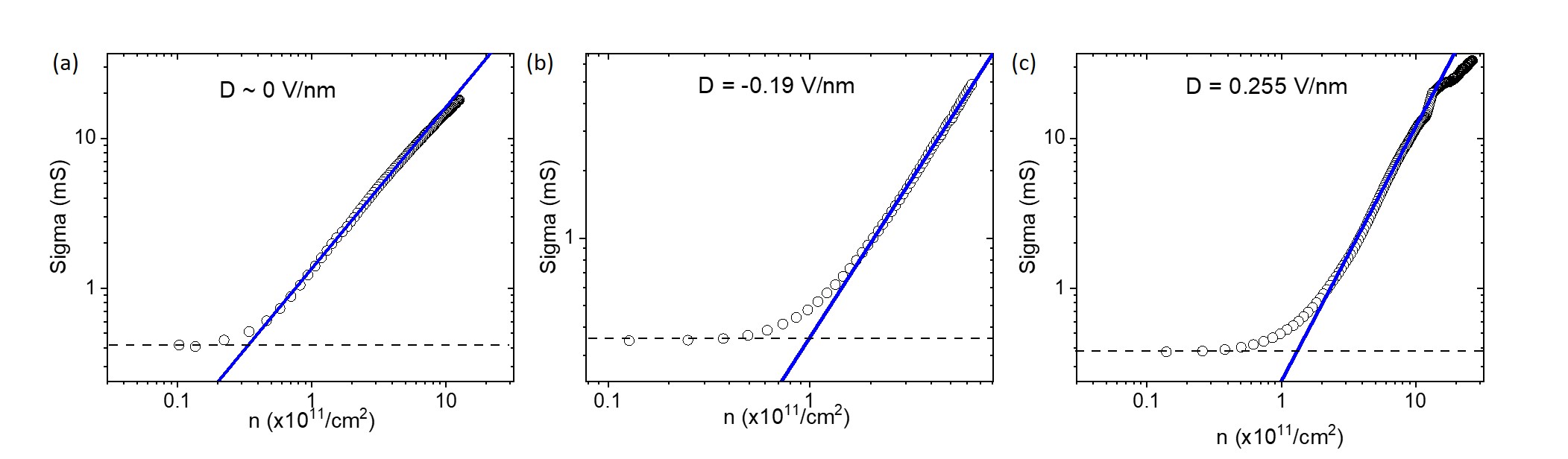} 
\caption{Estimation of charge inhomogeneity in graphene/monolayer WSe$_2$ control device. Double-logarithmic plot of $\sigma$ versus $n$ at  (a) D = 0 V/nm (b) D = -0.19 V/nm and (c) D = 0.255 V/nm respectively. The solid blue lines are a fit to the curve, and the minimum conductivity at the Dirac point is indicated by a dashed horizontal line.}
\label{S11}
\end{figure}

\subsection{C. Calculation of capacitive amplification}
The total number density and the displacement field as an effect of the dual gates is given by:
\begin{equation}
    D(V/nm)=[C_{tg}(V_{tg}-V_{t0})-C_{bg}(V_{bg}-V_{b0})]/2\epsilon_0
\label{t1}    
\end{equation}
\begin{equation}
    n(cm^{-2})=[C_{tg}(V_{tg}-V_{t0})+C_{bg}(V_{bg}-V_{b0})]/e
\label{t2}
\end{equation}
 where, $C_{bg}$ and $C_{tg}$ are bottom and top gate geometric capacitance respectively, $V_{tg}$ and $V_{bg}$ are the applied top and bottom gate voltages and $V_0$ is the gate voltage corresponding to the intrinsic doping. Upon solving the Eq~\ref{t1} and Eq~\ref{t2}, it is possible to independently vary either $D$ or $n$ by keeping the other parameter constant.  On substituting n = 0 in Eq~\ref{t2}, we get,
\begin{equation}
    \frac{\Delta V_{bg}}{\Delta V_{tg}}~=~-\frac{\Delta C_{tg}}{\Delta C_{bg}}
\label{t3}
\end{equation}
Therefore, using the locus of the Dirac points, we can estimate the ratio of capacitance and subsequently the hBN thickness from the slope of $n = 0$ line. We observe three additional capacitance values among which the diagonal line corresponding to primary Dirac point which matches well with the geometrical capacitance value (slope -6.76). However, the other capacitance values that comes from the additional lines D2, D3 and D4 does not match with the geometrical capacitance ratio, and indicates a large back-gate capacitance (Slope -4.41, -2.52, -0.68). Thus, the observed emergent inhomogeneity indicates an additional capacitance larger than the global back-gate capacitance. 

We consider the capacitive amplification to arise from negative capacitance (NC) of the ferroelectric layer. The negative capacitance in ferroelectrics can be understood in terms of a positive feedback mechanism \cite{salahuddin2008use, mcguire2017sustained}. Suppose we have a (positive) capacitor $C_{ox}$ (per unit area) that sees a terminal voltage equal to the applied voltage $V$ plus a feedback voltage $\alpha Q$ proportional to the charge on the capacitor $Q$ (per unit area), such that,
\begin{equation}
   Q = C_{ox} V + \alpha Q
\end{equation}
Hence, the effective capacitance $C_{eff} = \frac{C_{ox}}{1-\alpha C_{ox}} > C_{ox}$. Now, to calculate the individual capacitances, we use the series circuit model \cite{tu2018ferroelectric} as shown in Fig. 4(c) of the main manuscript. 
\begin{equation}
   \frac{1}{C_{eff}} = \frac{1}{C_{ox}}+\frac{1}{C_{FE}}
\end{equation}
where $C_{FE}$ is the capacitance per unit area of the ferroelectric. Using $C_{ox}$ = 0.01 $\mu F/cm^2$ for back-gate capacitance, from D2 we calculate a capacitive enhancement of 1.53 times. So, $C_{eff}$ = 0.0153 $\mu F/cm^2$. Using the circuit model we calculate $C_{FE}$ = 0.029 $\mu F/cm^2$.

\bibliography{References.bib}
\newpage
\end{document}